\documentclass[11pt,letterpaper]{article}
\usepackage{arxiv}

\usepackage[pdftex]{graphicx}
\usepackage{graphicx}
\usepackage{rotate}
\usepackage{wrapfig}
\usepackage{caption}
\usepackage{amsmath}
\usepackage{graphicx}
\usepackage{indentfirst}
\usepackage{subfiles}
\usepackage[T1]{fontenc}
\usepackage[english]{babel}
\usepackage{times}
\usepackage{amsmath,amstext,amsbsy,amsopn,amsthm,amsfonts,amssymb}
\usepackage{mathtools}
\usepackage{url}
\usepackage{tabularx,booktabs}
\newcolumntype{C}{>{\centering\arraybackslash}X}
\usepackage{booktabs,array,colortbl}
\usepackage{lastpage}
\usepackage{multirow}
\usepackage{algorithm}
\usepackage{balance}
\usepackage{placeins}
\usepackage{array}
\usepackage{blindtext}
\usepackage{subfigure}
\usepackage{caption}
\usepackage{capt-of}
\usepackage{booktabs}
\usepackage[small]{titlesec}
\usepackage{paralist}
\usepackage{verbatim} 
\usepackage[usenames,dvipsnames,svgnames]{xcolor}
\usepackage{algorithmic}
\usepackage{fancyhdr}
\usepackage{wrapfig}
\begin{document}

\title{Classification of Upper Arm Movements from EEG signals using Machine Learning with ICA Analysis}

\author{
	\begin{tabular}{ccc}
		Pranali Kokate & Sidharth Pancholi & Amit M. Joshi \\
		Electronics \& Commu. Eng. & Electronics \& Commu. Eng. & Electronics \& Commu. Eng.   \\
		MNIT, Jaipur, India. & 	MNIT, Jaipur, India. & MNIT, Jaipur, India. \\
		2019PEB5431@mnit.ac.in  & 
		sid.2592@gmail.com
		& amjoshi.ece@mnit.ac.in
	\end{tabular}	
}

\maketitle

\cfoot{Page -- \thepage-of-\pageref{LastPage}}
\begin{abstract}
The Brain-Computer Interface system is a profoundly developing area of experimentation for Motor activities which plays vital role in decoding cognitive activities. Classification of Cognitive-Motor Imagery activities from EEG signals is a critical task. Hence proposed a unique algorithm for classifying left/right-hand movements by utilizing Multi-layer Perceptron Neural Network. Handcrafted statistical Time domain and Power spectral density frequency domain features were extracted and obtained a combined accuracy of 96.02\%. Results were compared with the deep learning framework. In addition to accuracy, Precision, F1-Score, and recall was considered as the performance metrics. The intervention of unwanted signals contaminates the EEG signals which influence the performance of the algorithm. Therefore, a novel approach was approached to remove the artifacts using Independent Components Analysis which boosted the performance.Following the selection of appropriate feature vectors that provided acceptable accuracy. The same method was used on all nine subjects. As a result, intra-subject accuracy was obtained for 9 subjects 94.72\%. The results show that the proposed approach would be useful to classify the upper limb movements accurately.
\end{abstract}

%
\section{Introduction}
Neurotechnologists are increasingly developing methods and means to integrate brains and machines, enabling a myriad of real-time applications. The most significant of them are brain-computer interfaces (BCIs), which can provide alternative modes of communication and control of external equipment even in severe situations of impairment. Presently, there has been extensively research work is going in the deployment of external instrument control application platforms using real-time BCIs. EEG~\cite{zhang2019classification} is now the most accepted brain signal for practical execution of BCIs because of its long-lasting usability, high temporal resolution, ease of installation, and low prices. Human-brain interface is advanced to the point that it can do activities and achieve functional goals which were previously unlikely to be performed. In this field, research is developing in all aspects, be it the large scope of EEG-based BCI~\cite{ABDULKADER2015213} controls, the technological depth involved, and its usability for impaired people and the rest of the population.

\begin{figure}[htbp]
    \centering
    \includegraphics[width=4.5in]{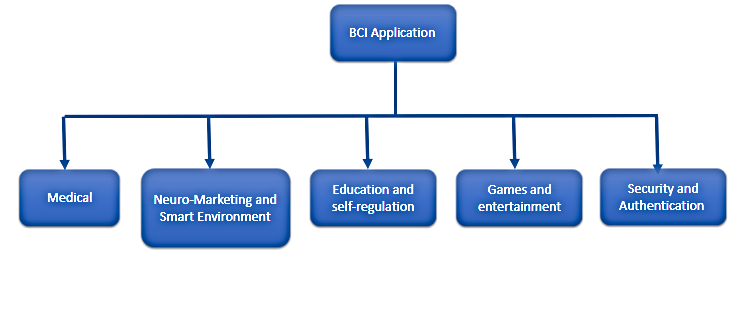}
    \caption{Various BCI-applications in different areas.}
    \label{bci-app}
\end{figure}

In the field of brain controlled applications, the publication of research has not only addressed medical requirements, but also addressed the demands of healthy users in their everyday lives. As briefed in Figure  \ref{bci-app}, the current growing BCI research comprises user-friendly EEG headsets and EEG signal analytics which have allowed BCI applications to be expanded to encompass automation, entertainment, emotion detection, medical automation diagnostics, neurorehabilitation and more. Every day, new and creative scientific advancements are being investigated in order to improve the lives of millions of people all over the world. Many researchers have developed EEG-based BCI systems to overcome the problem. An Electroencephalogram is created by collecting and storing electrical signals taken from the brain. EEG assists in deducing useful information about brain activities.

In general, some BCI techniques depend upon peoples' capacity to acquire command over their own activity in the brain through biofeedback, whilst other use classification algorithms that identify EEG sequence associated to certain voluntary objectives. The first experiments to allow individuals using feedback through their own activity in the brain were made in the 1960s. The technology allows people to develop voluntary control of their brain patterns. It has been observed that following training by EEG biofeedback, volunteers can identify their own alpha and mu frequency bands.
As a result, BCI research is interdisciplinary in nature, incorporating numerous domains such as brain physiology, cognitive neuroscience, electrical equipment, data transmission, data processing, pattern classification, artificial intelligence \cite{pancholi2019improved}, and so on. BCI system helps to decode the brain instructions and thus enabling the control of artificial arms particularly for the people enduring brain stroke, spinal injury, motor disabilities, and other injuries which lead to lost feedback mechanism between the brain and other parts of the body muscles. Furthermore, a BCI uses brain impulses to translate EEG signals into instructions for controlling external equipment such as  desktops, smartphones, wheelchairs, and prostheses.

\subsection{Motivation}
Amputation of the upper arm results in severe impairment \cite{pancholi2019electromyography}.
According to Dinesh Mohan's report “Amputee in India," an estimated 4,24,000 persons in India had arm amputations at or above the wrist.
The large percentage of arm amputations are caused due to trauma, such as workplace mishap or car accidents. People who have suffered a spinal cord injury, upper body paralysis, or a stroke lose the communication between the brain and the muscles tissues leading to arm dis-mobility.
 Today, BCI system has the opportunity to profoundly improve the quality of life of those with upper body disabilities.
 
 To resolve the problem many have developed an EEG-based BCI system to accomplish reliable performance with active hand-crafted feature extraction of EEG signal, but its clinical and cost consequences remain insufficient. Many scholars have attempted to improve accuracy in recent years by employing a variety of strategies. Since EEG recordings consist of a considerable quantity of data, signal interpretation and processing are critical. The mathematical modeling of EEG signals, as well as the extensive use of machine learning algorithms, are critical in the processing of this data and decision building. The paper covers the basic knowledge with cutting-edge approach to design a unique algorithm for the categorization of upper limb movements.

\subsection{Current Challenges}
The interaction among EEG sequence and computer intentions is a machine-learning challenge because the model must learn how to understand a specified EEG sequence. As with other learning issues, a training process is required to tackle this issue, wherein the participant is required to perform pre-specified mental processes (MAs) and a ML model is now in command of retrieving the affiliated EEG sequence. After the training process is complete, the participant must be able to monitor the machine's activities with his thoughts.
There are various issues with EEG-based BCI devices. One such issue is while recording signals. They are vulnerable to noise and are frequently polluted by artifacts. Similarly, EEG collection requires a costly setup, and when recording, they are interfaced with a variety of signals such as ocular signals, myoelectric signals, power line signals, and so on. The removal of them is necessary and requires a proper understanding of the signal. Feature extraction is difficult for motor Imagery tasks. There are rapid fluctuations in EEG data, and validation is a difficult challenge. Understanding and decoding of EEG signal is a critical task.

\subsection{Research Contributions}
The main research contributin have been derived as follows:
\begin{itemize}
    \item “EEG-based BCI system"\cite{zhang2019classification} is developed for upper Limb Amputees, stroke patients, and persons enduring motor disabilities. The performance has been validated on publicly available data set having motor imagery and motor execution areas. The identification and extraction of relevant signals from the raw EEG data is achieved with efficient methodology. 
    \item The effective EEG sensor configurations for various BCI applications has been selected from non-invasive  techniques. Moreover, the brain structure has been studied for effective electrode selection.
    \item  The efficient pre-processing methodology has been adapted for removing artifacts and cleaning signals.
    \item  The novel classification based Machine Learning model has been proposed for precise activities identification.
    \item The deep learning-based framework is designed for an EEG-based BCI system.
\end{itemize}

\subsection{Study of Hand Movements based on EEG and EMG approach}
Last 20 years, the advancement of brain–computer interfaces (BCI) has permitted interaction or command over external technology such as computers and artificial limbs using the electrical signals of the human neurological system.
In a typical BCI setup, the user is directed to visualise movement of various body parts (for example, right hand or leg motions), and the machine develops to detect distinct patterns of the synchronously collected EEG signals. The majority of the signal processing research for the advancement of EEG-based BCI systems has been designed to extract highly advanced features from the EEGs and refining/reducing the parameter dimensionality. 
Electromyography (EMG) data have primarily been employed in both rehabilitation and the design of prosthesis and haptic devices \cite{Pancholi2019EMBC,pancholi2020advanced}. As a result, EMG remains the most widely used and successful means of linking humans to machines. BCI (also known as brain–machine interfacing (BMI)) is a road-map that falls under the umbrella of human–computer interface (HCI), which connects cognition to actions.

Individuals with serious neuromuscular problems, like brain stem damage, amyotrophic lateral sclerosis, stroke, and cerebral palsy, may benefit from BCI systems based on EEG.  The EEG signal recording for an upper limb task has been examined in detailed manner~\cite{jeong2020brain}. Electrodes are implanted on the subject's head in considerable format to record. To ensure effective recording, the impedance between the electrode plates and the scalp is reduced. In 3D space, the person conducted center-out right-handed reaching action and images (left, right, forward, backward, up, and down). The participants sat on chairs in front of a desk. A display monitor was set roughly 60 cm away from the patient on the desk to provide a convenient display distance. Visual signals in the form of a clue for rest and relevant navigational signals for reaching activities were displayed. The experiment was divided into two sessions: motor execution and visualisation. The individuals performed the motor movement, such as center-out arm extending in one of the directions, during the motion execution. During the imagery session, the participants were just required to perform motor imagery.

The P300 wave is a form of Event-Related Potential (ERP) that is thought to be prominent in making decision, which normally occurs between 250 and 500 milliseconds after the start of a visual clue. Simple Reaction Time (SRT), similar to the P300 wave, reflects the time lag between visual clue and reaction. Following the reception of a visual input in the visual cortex, visual data is conveyed through two distinct channels, namely the ventral and the dorsal streams. The ventral stream finally reaches the temporal cortex, which is responsible for picture identification. With the aid of relevant memory, the visual input is therefore utilized to form the connection between experiment commands and completing L/R arm movements.

The acquisition of myoelectrical signals are very important~\cite{jamal2012signal}.   In most situations, two sensing surfaces (or EMG electrodes) are bipolarly implanted on the skin. To achieve the expected potential signal, the EMG electrode should be positioned correctly, and its direction across the muscle is critical \cite{pancholi2018portable}. Surface EMG electrodes should be positioned along of the muscle's lengthwise centerline, in between motor unit and the tendon insertion \cite{pancholi2019time}. The spacing between the electrodes or detecting surfaces should be no more than 1-2 cm. The electrodes' longitudinal axis (which travels across both sensing surfaces) should be proportional to the extent of fibers. Individuals who have had their left hand amputated should glance at their right hand to replicate the movements.
Despite the ease of acquisition of EMG Signals for Upper limb movement \cite{Pancholi2021iULP}, EEG-based approaches may identify brain activity modulations that correspond with visual inputs, gaze angle, intentional intents, and cognitive processes. These benefits prompted the creation of numerous types of EEG-based systems, which varied based on the cortical regions monitored, the extracted EEG characteristics, and the sensory methodology delivering input to participants.

\subsection{Techniques in BCI: Investigation of EEG and its Counterparts}
Differently abled individuals interact with the outside world (humans and assisting technologies) via brain-computer interface technology, which translates human neural activity into command signals for neurofeedback and monitoring purposes. Bioelectrical recordings, which are generated by analyzing electrophysiological or hemodynamic responses, are important clinical methods for the construction of the both BCIs. The electrical and hemodynamic neural activities are collected directly from the cerebral cortex, either invasively, partially invasively, or noninvasively, to build BCI systems. Invasive and partially invasive BCI signal acquisition requires the neurosurgical implantation of micro-electrodes inside the cerebral cortex or underneath the scalp or the surface of the cerebrum. The following subsections discusses about the many forms of invasive and non-invasive neuroimaging methods.

\subsubsection{Invasive BCI Techniques}
\begin{enumerate}
    \item \textbf{Electrocorticography (ECoG)}\\
    ECoG measures brain responses by measuring the electrical activity of neurons on the cortex's surface following micro-electrode array implantation. Because signals are recorded from more proximal and concentrated areas of the brain, the resulting signals have higher temporal-spatial resolution, signal amplitude, and spectral bandwidth than noninvasive EEG data. These acquired signals have high-amplitude signal, less noisy and absent of muscular motion artifacts. ECoG signal collection has already been used to build a variety of BCI systems. By studying P300 ERPs and evaluating alpha, beta, or gamma sub-bands, a team of experts was able to correctly categorise separate voluntary-involuntary motor motions, cursor control, or operate robotic arm.
    
    \item \textbf{Intracortical Neuron Recording}\\
    The electrical activity of neurons is acquired by the intracortical neuron recording using the single or array of sensors placed in the cortex, i.e. extremely near to the signal origin. It is an invasive approach which may produce brain signals with extremely high temporal and spatial resolution. However, due to higher brain tissue resistance during signal transmission from deep-implanted electrode to recording equipment, it could suffer from neuron death or the incorporation of a noise component.
\end{enumerate}
\subsubsection{Non-Invasive BCI Techniques}
Noninvasive BCI data collection measures,the brain activity using exterior surface electrodes deployed over the scalp of the subject. EEG, MEG, fMRI, fNIRS, PET, SPECT, and a variety of other noninvasive neuroimaging methods are employed in BCI research.

\begin{enumerate}
    \item \textbf{Magnetoencephalography (MEG)}\vspace{8pt}\\
    MEG is a noninvasive neurological recording technique that captures the magnetic activities produced by natural sources like the electrical impulses of neurons in the brain. Magnetic fields are generated by the passage of intra-cellular currents, which causes magnetic induction. A precise superconducting quantum interference device (SQUID) is used to detect and record magnetic signals in MEG. These sensors are capable of capturing mynute magnetic fluctuations caused by neuronal rhythms in the brain. MEG captures magnetic fields that are less contaminated and deteriorated than EEG electrodes that records electric fields. Despite these benefits and its close association with EEG, MEG is not a popular brain imaging mode for BCI design. The main drawback is its non-portability, cost, and complex acquisition mechanism. Furthermore, MEG signal analysis approaches are primarily concerned with the characterisation of stimulus-induced neural activity rather than motor activities.
    \item \textbf{Functional Magnetic Resonance Imaging}\vspace{8pt}\\
    fMRI is yet another non-invasive technique of brain scanning method that uses electromagnetic fields to detect and record. The entire procedure is based on the idea that engagement of any brain area for specific activity is connected with increased blood flow in the region of interest (ROI). As a result, fMRI is utilised to assess oxygen levels in blood flow across active brain regions with changes in local cerebral blood flow and oxygen levels in blood flow. fMRI has a relatively poor temporal resolution of roughly 1–2 sec. It also addresses the issue of 3–6 sec physiological delay. As a result, fMRI-BCIs are unsuitable for high-speed BCI interactions, which are required in practically all nonclinical applications. Furthermore, the pricey, non-portable, and heavy MRI scanners are prone to electromagnetic disturbance and dynamic artifacts.
    \item\textbf{Functional Near-Infrared Spectroscopy (fNIRS)}\vspace{8pt}\\
    The fNIRS is a noninvasive brain wave collection method that uses near-infrared light to quantify alterations in localised cerebral metabolism while performing particular neural activity. It is based on optical spectroscopy. To measure blood flow, near-infrared light can permeate the brain skull up to a depth of about 1–3 cm beneath the cerebrum surface. The fNIRS measuring instrument is less costly and transportable, however it has limited brain scans features. This optical imaging approach cannot reach the inner cortical areas. It also has a delay in receiving hemo-dynamic responses corresponding to specific neural activity. Motion artifacts caused by head movements and obstruction caused by hair may further degrade signal quality and thus the performance of designed BCI.
    \item\textbf{Electroencephalography (EEG)}\vspace{8pt}\\
    EEG, which was first reported in 1929 by a German psychiatrist named Berger H. Uber, includes the capturing of brain electrical impulses by deploying sensors across various scalp areas. An electrical magnitude (microvolts) vs. period records are obtained by nonpolarized and interference-free sensors. They are all positioned over the scalp in such a way that they must acquire brain response from all of the cerebrum's regions, including frontal, temporal, parietal, and occipital. Electro-headcaps with electrodes mounted on them are frequently used to obtain multi-channel EEG recordings with a huge number of sensor pairs. EEG is a prominent noninvasive neurodiagnostic technique for monitoring neural activity with high temporal resolution, or the capacity to detect changes in neural activity over a short period of time. Conversely, because the amplitude of neural electrical impulses deteriorates during transmission from inbuilt cortical areas to scalp areas, it performs poorly in terms of spatial resolution and SNR. BCI scientists are attempting to lower the number of sensors necessary to collect neural  activity yet maintaining a high SNR.
\end{enumerate}
Additionally, ECoG and intra-cortical recordings depend on the placement of micro-electrode grids underneath the cerebrum and inside the cortical areas of the brain, respectively. Magneto-encephalography measures magnetic fields associated with particular brain actions. While fMRI measures minor fluctuations in blood oxygen levels caused by associated neural actions. Likewise, fNIRS assesses cellular function in the cortical region of the brain by measuring the hemo-dynamic optical response. Researchers have successfully used all of these neuro-imaging modes to construct BCIs. ECoG and intracortical recordings, on the other hand, are intrusive techniques of signal collection that need embedded electrodes.
The dependability, safety in surgical operations, consistency of function, and ease of these approaches are key considerations. MEG and fMRI are linked with extremely strong magnetic fields and are limited by the highly costly and immobility of big acquisition devices. In addition, the brain waves are collected with a limited temporal resolution. Due to these constrains, these techniques are unsuitable for high-performance main BCI control systems when evaluated to EEG, a noninvasive neuro-imaging technique with excellent temporal resolution.
\subsection{Organization of Paper}
\begin{figure}[htbp]
    \centering
    \includegraphics[width=7in]{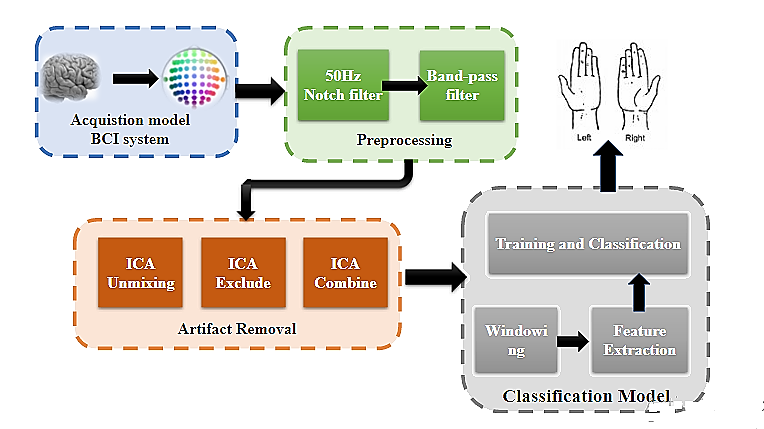}
    \caption{EEG Based BCI architecture}
    \label{class}
\end{figure}

This study has aided in the improvement of each main component involved in Figure \ref{class} including Proper channel selection, pre-processing, Artifact removal, and new categorization method for motor activities.
Python-mne\cite{mne_ica} package was used to avail the knowledge regarding the Brain signals. Also explored the EEGLAB toolkit available in MATLAB to understand the concept of source localization. 
 
\begin{itemize}
    \item Section 2 discusses the various state-of-the-art for EEG signal based on BCI. This chapter highlights the various forms of machine learning architecture and data sets for motor imagery and execution class. It also cover all the prior methods for EEG data processing, signal segmentation, and feature extraction. Various architectures are studied for deep learning analysis.
    
    \item Section 3 elaborates background theory related proposed work. It talked about frequency waves related with the brain, as well as various forms of noises. The mathematics underlying ICA-Independent component analysis has been also explored.
    
    \item Section 4 emphasis on a publicly available dataset for motor activities. The sectiion also covers: 1) the selection of channels is a key task that must be done correctly. 2) Pre processing is required prior to artefact removal, which is covered in the length of this chapter. 3) Feature extraction is an essential step. 4) The proposed model as well as deep learning architectures are explained.
    
    \item Section 5 describes the results along with comparison with state of art work. The conclusion and future work is discussed in Section 6. 
    
    
\end{itemize}

\section{Literature Review}
In this section, various available data sets related to motor imagery and execution activities are discussed along with several features associated with EEG signals. Further, It also covers the different Pre-processing technique and Various classification algorithms from machine learning to deep learning \cite{Sidharth2021DLPR}. 

The paper introduced an algorithm that classifies  for movement of left and right hands~\cite{7993477}. They considered 9 volunteers for their study. Each individual would go through a series of 30 trials of elbow and shoulder movement. The individuals were given visual stimuli to stimulate their imagination. The experiment lasted 8 seconds in total with stimulus was delivered at 1sec.  Pre-processing was carried using MATLAB software. An IIR filter with a ripple factor of 1 and a stop band attenuation value of 60 was adopted. The signal was then routed through a band pass filter with a frequency range of 8-30 Hz.  Only two channels associated with motor cortex and pre-motor cortex were considered. They extracted entropy and power spectral density and utilised SVM for classification, achieving an accuracy of 91.25 percent for four classes.

The paper \cite{mousapour2018novel} investigated the connectivity pattern between brain areas during MI task. From the study, they obtained features such as clustering coefficient, Path Length, Efficiency, Modularity, Betweenness Centrality and Eigenvectors. They considered the publicly available data set called BCI Competition IV. The data set consists of 9 test volunteers who were instructed to visualize movement of their hands, feet, and tongue. The mental task was carried out with the help of visual cue. Signals were pre-processed using techniques such as a 50 Hz filter followed by a bandpass filter.  These attributes were transformed into an image for further classification by repeating the values. The 2D image was further processed using CNN and the model provided an accuracy of 96.69\% for binary classification and multi class classification had accuracy of 86\%.

In paper \cite{8298770}, they proposed a model to categorize right and left hand imagery movement. There were a total of 80 sets in the data set. They employed 10 channels for their experiment, the visual stimulus was activated every 2 seconds, and the experiment lasted 10 seconds. ICA was used to eliminate extraneous signals. Following data cleaning, common spatial pattern characteristics were retrieved. The SVM optimization strategy based on the Genetic Algorithm was used and  achieved an average accuracy of 95\%.

The non-linear feature extraction technique related to Higher Order Spectra (HOS) was retrieved in the paper ~\cite{salwa2018classification}. They conducted an experiment with single participant performing random left and right imagined hand and leg movement using a visual stimulus. For acquisition, 18 channels were considered for this experimentation. They calculated the Maximum and Entropy value for both magnitude and phase values of HOS spectra. SVM was utilized as a classifier, and average multi-class accuracy was 79.14\%.

In paper \cite{benzy2019classification}, they examined the EEG signals associated to the MI task for hand movement (controlling hand) in the left and right motions. The PLV (Phase Lock Value) characteristics of the most important EEG sensor combinations in the specified frequency range were retrieved as discriminative characteristics for categorization of hand movement orientations. Data was collected from 12 participants using a 64-channel BrainVision acquisition device. To reduce high frequency noise, these signals are band passed (0.05-45) Hz. The Naive Bayes Classifier was used for binary classification. The outcomes are addressed in terms of the Subject-specific frequency band and the electrode pair. Subject one achieved the best accuracy of 88.2\% in the 4-8 Hz range.

In paper~\cite{chowdhury2019processing}, they used a B-Alert EEG headset to acquire the data. The electrodes were set in a 10-20 channel electrode placement method, and individuals were asked to visualize hand motions. For this research, three participants were considered. They were instructed to perform three activities utilizing a visual stimulus: rest, right-hand movement, and left-hand movement. After recording of EEG data they are further processed to eliminate power line artifacts, the recorded EEG data is cleaned using a notch filter in the frequency range 48 Hz to 52 Hz. The filtered data is then cross-correlated in order to obtain the necessary cross-correlograms (CC). They derived six statistical variables from each result cross-correlogram sequence, including mean, median, mode, standard deviation, minimum, and maximum~\cite{chowdhury2019processing}. SVM was employed as a classifier. They achieved a cross-validation average accuracy of 78.75\%.

EEG sub-bands are connected with different brain states. The most efficient frequency ranges representing brain processes associated to MI tasks were discovered by trial and error in this study~\cite{9219507}. The DWT approach was used to split EEG signals into sub-bands and transform it into images simultaneouslyr. For their analysis, they employed a publicly available dataset named BCI Competition 2003. They trained the obtained DWT images using CNN model and achieved an accuracy of 90.38 percent.


The paper \cite{hossain2020detection} discussed BCI activities that are mostly linked to motor imagery activities. They utilised a publicly available data set of 13 participants and acquired EEG signals with a sample rate of 512Hz using 64 channels. They concentrated on the actions of the right and left hands. They extracted seventeen feature vectors from sensor data across a 700-second span. The characteristics are mean, median, standard deviation, mean absolute deviation, Signal Interquartile Range, Quantile25, Quantile75 Peak2peak value, RMS Value, crest Factor, Shape Factor, Impulse Factor, Margin Factor, Signal Energy, sample skewness, kurtosis and Entropy \cite{hossain2020detection}. They used the minimum redundancy maximum relevance (MRMR) technique to rank the features. They used SVM as a classification algorithm and obtained an F1-score of 68.69 \%. 

\subsection{Research Gap}
The research gaps from literature have been observed as follows:
\begin{itemize}
    \item There are EEG signal acquisition devices available in the market, but they are quite expensive and complex to handle. The cost rises as the number of channels increases. The number of channels is important since it influences the categorization outcome.
    
    \item There are few studies in the literature that discuss the proper selection of electrodes, and even fewer research discuss regarding the source of information flow as per electrodes. Because the proper selection of electrode may affect the accuracy due to loss of vital information. The study must be done on selection and information flow through electrode.  

\item The fundamental disadvantage of EEG signals is that they are vulnerable to both external and other biological influences. There must be a thorough discussion as well as some conventional ways for removing the tainted noises. Validation of the applied procedures is required.

\item For the EEG-based categorization of motor activity, there is a significant gap between offline processing and online validation. The current work technology has been verified in the laboratory only, which has the requirement of complicated system.
\end{itemize}

\section{Background Theory}
Several types of waves associated with the brain have been explored, followed by an understanding of numerous forms of artifacts and their classification. Learned about the ICA methodology, which aids in the removal of artifacts. Later, ideas were comprehended  related to machine learning and deep learning. Finally, investigated the artificial generation of EEG signals with GANs and the usage of Autoencoders on CNN architecture.

\subsection{Brain Rhythms}
\begin{figure}[htbp]
\centering
\includegraphics[width=4in]{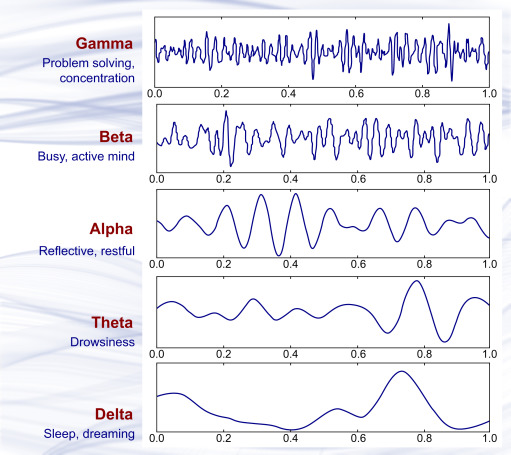}
\caption{Frequency band associated with Brain rhythms}
\label{freq}
\end{figure}

According to the \cite{sanei2013adaptive}, The amplitudes and frequencies of brain impulses shift from one state of a human to the next, such as sleep and awareness. Brain Rhythms are related with five primary brain waves.
These frequency bands, which have both low and high frequencies, are known as: \\
\begin{math}
1. Alpha (\alpha)\\
2. Theta (\theta)\\
3. Beta (\beta)\\
4. Delta (\delta)\\
5. Gamma (\gamma)
\end{math}
\begin{itemize}
\item \textbf{Delta waves} have frequencies ranging from 0.5 to 4 Hz. These waves are mostly associated with deep sleep. In general, it may be confused by artifact signals caused due large muscles of the neck and jaw.
\item \textbf{Theta waves} line within the range of 4 to 7.5 Hz. These waves are generally associated with consciousness slips towards drowsiness, deep meditation. They play an important role in infancy and childhood. These waves are generally studied while examining maturation and emotional studies.
\item \textbf{Alpha waves} are present in the posterior part of the head, often over the occipital area of the brain. These sinusoidal waves have a frequency range of 8-13 Hz. These waves are commonly connected with relaxed awareness and lack of attention. When the subjects' eyes are closed, they form this pattern. These have a greater amplitude over the occipital region and an amplitude less than 50 $\mu V.$
\item \textbf{Beta wave's} electrical activity lies within the range of 14-26 Hz.  This brain rhythm is often linked to active thinking, active attention, and problem solving. An individual in a panic state has a higher amplitude. These are more common in the frontal and central areas. The beta wave's amplitude is less than 30 $\mu V.$
\item \textbf{Gamma wave's} frequencies are above 30 Hz mainly up to 45 Hz. These are known as rapid beta waves. The gamma wave band has also been shown to be an excellent indicator of brain event-related synchronisation (ERS) and may be used to indicate right and left index movement. Figure \ref{freq} shows the different frequency bands associated with brain.
\end{itemize}

\subsection{Artifacts Classification}

During the neurological acquisition process, some peripheral electrical and muscle patterns are taken up in addition to neuronal patterns. These artifacts are repeated temporal and spatial patterns that come from non-cerebral areas and detract from the gathered neural responses. The existence of artifacts may worse the classification process's performance. As a result, identifying and removing artifacts is necessary prior to further signal processing.\vspace{8pt}\\
The classification of EEG artifacts which causes interference in the signals  has been discussed~ \cite{bansal2019eeg}.
EEG artifacts are generally categorized as:
\begin{enumerate}
    \item \textbf{Physiological}: 
Artifacts created by peripheral electrical and muscle activity.
    \item \textbf{Non-physiological} 
Artifacts generated by interference from electromagnetic sources.
\end{enumerate}
\vspace{16pt}
\textbf{Physiological Artifacts:}\\
EEG data collected from certain brain areas are susceptible to artifacts induced by bio-electrical activity such as unintentional eye movement or eye blink known as EoG, electromyographic (EMG) muscle movements and muscle contractions electrocardiograph (ECG) artifacts, pulse artifacts.\vspace{16pt}\\
\textbf{Eye Movement Artifacts}:\\
Whenever the eyes blink, it generates potential due to charge in variations. 
As a significant potential difference is formed between the positively charged cornea and negatively charged retina, eye blinking causes very high-intensity brain patterns. These physiological artifacts, such as eye movements, are particularly prominent in the cerebrum's frontal lobes.\vspace{16pt}\\
\textbf{ECG artifacts}:\\
 ECG artifacts are mostly induced by the powerful muscle dipoles induced by atria and ventricular contraction and expansion. These rhythmic patterns interfere with brain waves and can have a magnitude up to ten times that of EEG waves. This artifact may be identified by its fixed wave pattern, which consists of the P-QRS-T complex. \vspace{16pt}\\
 \textbf{Pulse artifacts}:\\
 When the electrode is mounted over a pulsing blood vessel, pulse artifacts occur. The artefact has a consistent wave pattern and is immediately identifiable.
Moving muscles adjacent to the scalp, such as the forehead and neck muscles, causes muscular movement artifacts to appear across the temporal and frontal scalp areas. \vspace{16pt}\\
\textbf{Non-physiological Artifacts:}\\
External electromagnetic disturbances, such as power-line interference with frequencies of 50/60 Hz, variations in electrode resistance, cable movements, and a low-intensity battery of the acquisition unit, mostly affect EEG signals.\vspace{16pt}\\
\textbf{Power-line interface:}\\
Main power lines cause power-line interference due to alternating current sources such as electronic gadgets.\vspace{16pt}\\

\begin{figure}[!h]
\centering
\includegraphics[width=5in]{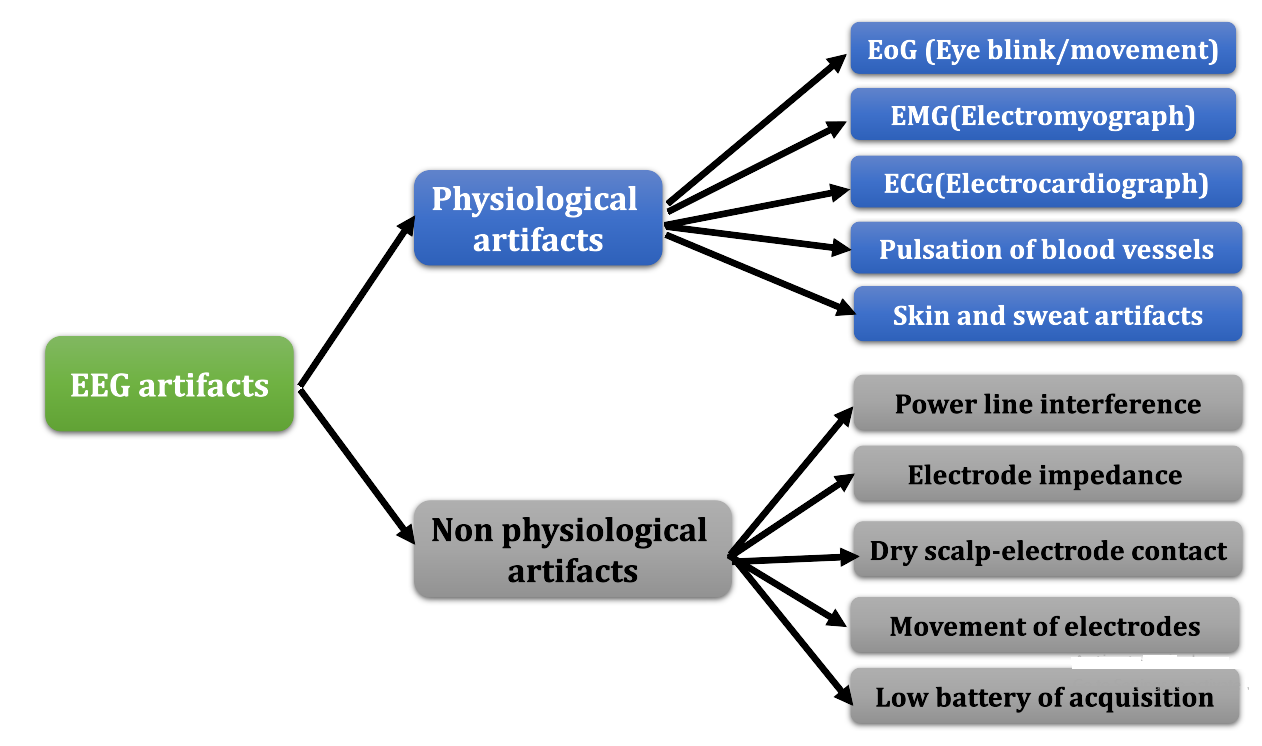}
\caption{Different types of artifacts\cite{bansal2019eeg}}
\label{artifacts}
\end{figure}

\textbf{Demodulation Artifacts:}\\
High-frequency waves, such as radio/microwaves, produce artifacts at frequencies greater than 10 Hz. These high-frequency waves interfere with low-frequency EEG pulses during demodulation. If the person rubs their feet, shoes against the floor, or their hands against each other while recording, the ground potential may rise.

Because of lateral movement between electrodes, friction occurs, resulting in tribo-electric charges on the electrode surface. When electrodes with inadequate resistance connections are employed, these artifacts predominate. Internal sources may lead to the formation of shot, flicker, or thermal sounds, which can cause fluctuations in acquired brain potentials~\cite{bansal2019eeg}. The representation of various artifacts are presented in Figure \ref{artifacts}~\cite{de2009handbook}.

\subsubsection{Artifact Detection}
EEG artifact detection requires accurate analysis of incoming brain signals across certain time frames, as well as the use of amplitude outsets to identify the artifacts. For example, the length of muscular artifacts is shorter when compared to EEG patterns, but they are unpredictable in both time and amplitude.
However, identifying artifacts is critical for subsequent pre-processing of collected brain signals to extract significant brain activity information. 
Artifacts can be detected by carefully studying collected EEGs over a given time frame and using amplitude threshold methods~\cite{zhang2020arder}.

EEG artifacts detection is a crucial task and the different techniques are discussed briefly.
The paper used Haar basis discrete wavelet transform to separate ocular artifacts~\cite{zhang2020arder}. Furthermore, in the paper, they proposed MAD for Muscle artifacts detection, which calculated the proportion of energy in the 30-100 Hz frequency band to the whole EEG signal. While they estimated Kurtosis which is capturing the 'peak' state of the Transmission line Artifacts.

Next, The paper converted an EEG signal into a two-dimensional image and they used two classes first as pure EEG signals which are free from ocular artifacts and another class contaminated with ocular artifacts~\cite{mashhadi2020deep}. These images are given to the training model for classification. Here the judging parameter is the Mean Square Error (MSE). 

Further, The paper~\cite{sai2017automated} calculated the wavelet ICA of the EEG signal where
ICA decomposes the EEG signal into independent components. Then they calculated a set of features including Kurtosis, Variance, Shannon's entropy. These features were divided into training and testing datasets and provided to the SVM model. The result produced contaminated IC's which were further dropped the IC's and inverse ICA was applied on the EEG signals~\cite{sai2017automated,mashhadi2020deep}.

\subsection{Independent Component Analysis (ICA)}
EEG signals are the combination of different signals and Independent components analysis (ICA) is a method for evaluating independent source signals from a collection of recordings in which the source signals were combined in unfamiliar proportions. For example, it has been considered a common example of blind source separation \cite{4536072}. For this, Let's consider three musical instruments playing in the same room, and the three microphones recording the performance here it must noted that each microphone is accumulating sound of all three instruments, but at different levels.
One must somehow try to  "unmix" the recorded signals from three different microphones so that end up getting separated recordings, separating the sound from each instrument. 

The similar analogy was used to fits EEG/MEG analysis: while recording EEG signals gets affected by different artifacts link blinks, heartbeats, activity in different areas of the brain. It must be noted that these source signals must be statistically independent and non-gaussian so that it is possible to separate the source signals using ICA, and after separating must be possible to re-combine the sensor signals after removing the contaminated part. The data is captured as linear sums of separate components derived from spatially specified sources.
The goal here is to generate N independent source signals from N linearly mixed acquired signals recorded by the N electrodes as per Equation (\ref{unmixing_equation})~\cite{7319296}.

\begin{equation}
\label{unmixing_equation}
   x = As 
\end{equation}

 Here \textbf{A} is unknown mixing matrix. Here the critical issue in ICA is to find the  un-mixing matrix \textbf{W}. Which is nothing but the pseudo-inverse $A^{-1}$ and s are the sources which are independent to each other statistically.

\begin{equation}
\label{inverseequation}
   y = Wx 
\end{equation}
Specific methods across the time series determine the independence of the basis vectors \textbf{y}. In order to solve Equation (\ref{inverseequation}), It is assumed  that the mixing process is linear. \textbf{W} may be thought of as a spatial filter matrix that inverts the mixing matrix linearly. The maximising of non-Gaussianity is used to determine independent components \textbf{Wx}\cite{4536072}.

There are various types of  ICA algorithms \cite{4536072} some of which are discussed here: fastica, Picard, and infomax. FastICA and Infomax are both inequitably extensive use; Picard is the latest (2017) algorithm that is supposed to converge faster than FastICA and Infomax and is more robust than other algorithms.

\subsection{Fundamental Concepts}
\subsubsection{Neural Network}
A Neural Network is made up of a large number of artificial neurons that are connected using a specific network architecture \cite{jain2020iglu}. The neural network's primary goal is to convert inputs into meaningful outputs \cite{joshi2020iglu2.0}. One input layer, one output layer, and a number of hidden layers make up a Neural Network architecture~\cite{kansara2018visual,8922820}.

\begin{figure}[htbp]
\centering
\includegraphics[width=2.7in]{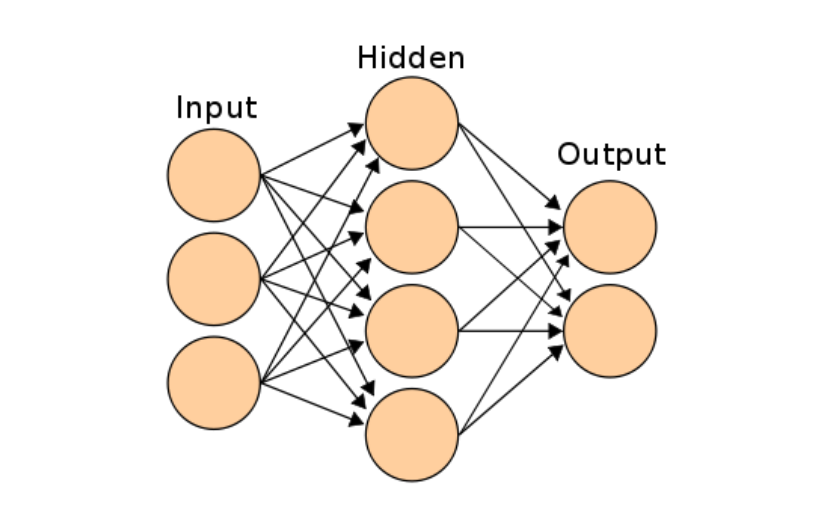}
\caption{Neural Network Architecture}
\label{Fig-1}
\end{figure}

\subsubsection{Convolution}
Convolution is a mathematical operation that is done on two input functions. It returns the point-wise product of the two input functions as an integral. With regard to the other function, it may be viewed as a modified version of the original functions~\cite{kansara2018visual}.
\begin{figure}[htbp]
\centering
\includegraphics[width=1.5in]{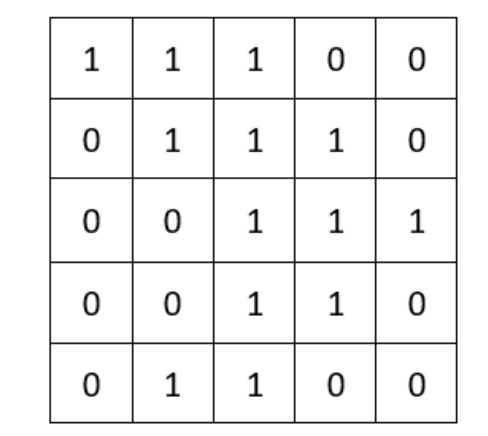}
\includegraphics[width=1.3in]{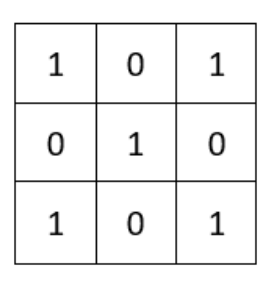}
\includegraphics[width=1.4in]{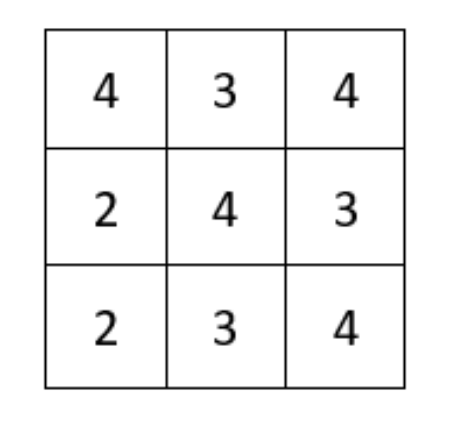}
\caption{Convolution Example}
\label{Fig-2}
\end{figure}

Consider the 5x5 matrix in Figure \ref{Fig-2}, with 0s or 1s as elements. As a filter, another 3x3 matrix, similarly made up of 0s and 1s, is utilised in the picture. These two matrices can be understood as the convolution function's two inputs. The convolved feature is obtained by conducting element-wise multiplication by sliding the filter over the original picture.

\subsubsection{Convolution Neural Networks}
Convolutional Neural Networks (CNNs) are stacked layers of convolutions with non-linear activation functions such as tanh or ReLU. Each node in the current layer is connected to every node in the next layer in a neural network. Convolutional filters, on the other hand, are employed in CNNs to slide over the nodes in the input layer and subsequently compute the output~\cite{kansara2018visual}. The output of each layer of the network is computed using different convolutional filters. When the network sees the training data, it learns which filters to employ automatically.\\
\begin{figure}[htbp]
\centering
\includegraphics[width=6.5in]{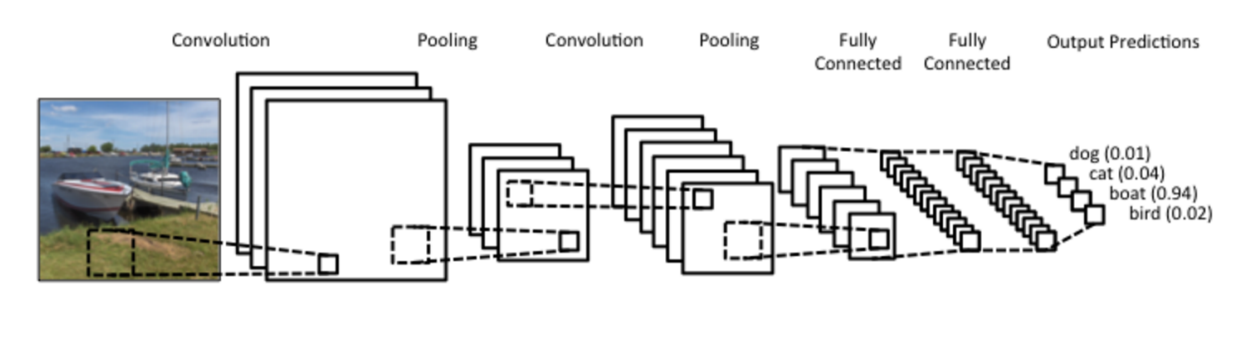}
\caption{Convolution Neural Network Example}
\label{Fig-3}
\end{figure}

\subsubsection{Recurrent Neural Networks}
Recurrent Neural Networks (RNN) are neural networks that execute the same job for each element of the input sequence. The RNN's output is determined by the current element as well as prior calculations. RNNs have a memory that stores all of the preceding data's information~\cite{kansara2018visual}.

\subsubsection{Long short-term memory}
The long short-term memory (LSTM) network is a sort of recurrent neural network that can learn and recognise long-term dependencies and is utilised as a block of hidden layers \cite{sharma2021dephnn}. The LSTM is a recurrent network in and of itself, as it contains recurrent connections identical to those seen in a traditional RNN. The gating mechanism in an LSTM block is made up of four parts: a cell, an input gate, an output gate, and a forget gate. The cell is in charge of remembering values over arbitrary time intervals, which is why LSTM uses the word memory. Each of the three gates can be compared to a traditional artificial neuron in a feed-forward neural network, in that they compute activation using a weighted sum activation function~\cite{kansara2018visual}.
\begin{figure}[htbp]
\centering
\includegraphics[width=3.5in]{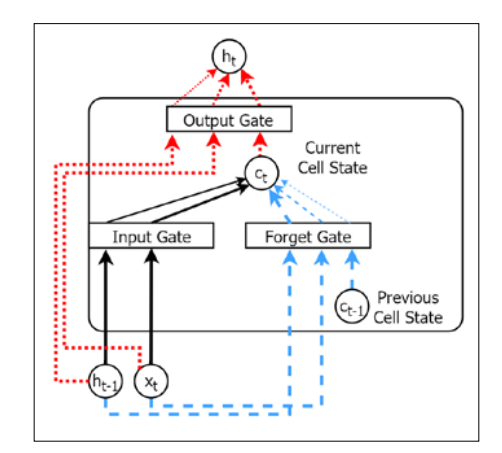}
\caption{LSTM Data Flow}
\label{Fig-4}
\end{figure}

LSTM are commonly used in sequence tasks to extract required information. Attention based LSTM are given priority since it will remember what it wants to remember. 

\subsubsection{Activation Functions}
An activation function is used in neural networks to evaluate the output of a specific node for a given input set of nodes from the preceding layer. There are several types of activation functions, both linear and non-linear~\cite{kansara2018visual}. Some of the most prevalent are listed below.\\
\textbf{Hyperbolic tangent or tanh:}\\
The hyperbolic tangential function is a non-linear activation function given by:
\begin{figure}[htbp]
\centering
\includegraphics[width=3in]{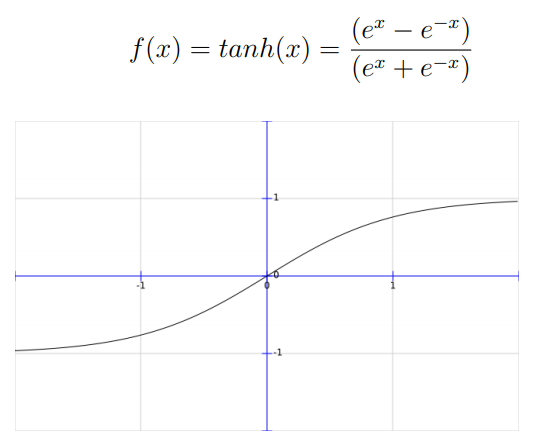}
\caption{Tanh function}
\label{Fig-5}
\end{figure}

Tanh is a widely used activation function. Its popularity stems from its high gradient, which makes it a superior choice for network back propagation. Tanh's first order derivative is provided by Equation (\ref{eq8}):
\begin{equation}\label{eq8}
    \begin{split}
        f'\left( x\right) =1-f\left( x\right) ^{2}
    \end{split}
\end{equation}
\textbf{Rectified Linear Unit or ReLU:}\\
The positive component of the input is defined as the ReLU activation function. It is expressed mathematically as:
\begin{equation}
    \begin{split}
        f\left( x\right) =0 \hspace{8pt} for\hspace{8pt} x <0
    \end{split}
\end{equation}
\begin{figure}[htbp]
\centering
\includegraphics[width=3.5in]{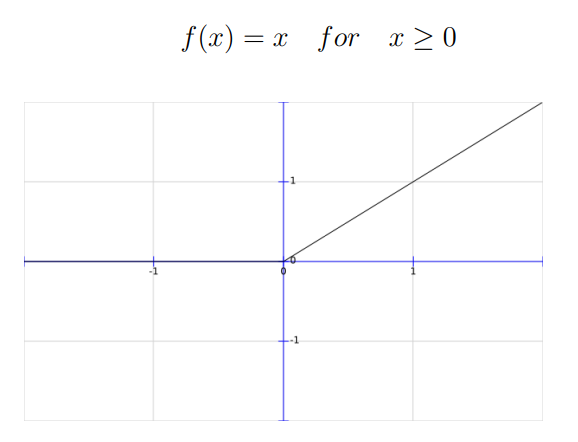}
\caption{ReLU function}
\label{Fig-6}
\end{figure}

Because it increased the performance of deep learning models, ReLU is also one of the most extensively utilised activation functions for deep neural networks.\vspace{8pt}\\
\textbf{Softmax:}\\
The Softmax activation function is mostly employed in the neural network's final output layer, which is employed for multi-class categorization. The softmax function returns a probability distribution across the number of target variables, with the class with the greatest probability value estimated as the final output~\cite{kansara2018visual}.

\subsubsection{Optimizer}\vspace{8pt}
\textbf{Adam Optimizer:}\\
Adam optimization is a stochastic gradient descent approach based on the adaptive estimate of first- and second-order moments. According to the study ~\cite{kingma2014adam}, the approach is computationally efficient, requires minimal memory, is insensitive to diagonal re scaling of gradients, and is well suited for big data/parameter issues.
Adam optimizer's characteristics are as follows~\cite{reddi2019convergence}:
\begin{enumerate}
    \item \textbf{learning\_rate:} A Tensor vector or a floating point value. 0.001 is the default value.
    \item \textbf{beta\_1:} A float value or a tensor of constant floats. The rate of exponential decline for first-moment estimations.
    \item \textbf{beta\_2:} A float value or a tensor of constant floats. The rate of exponential decline for 2nd moment estimations.
    \item \textbf{epsilon:} A small constant for numerical stability. 
\end{enumerate}

\subsection{Artifical EEG generation using Generative Adversarial Nets (GANs)}
BCI has always been a source of concern, significant data-related issues, including a lack of adequate data
as well as data corruption. The use of GANs~\cite{fahimi2019towards} for time-series data creation is a new and developing field that must first be evaluated for viability. Researchers explore the performance of GANs in producing artificial electroencephalogram (EEG) signals. The findings indicate that GAN-generated EEG signals have temporal, spectral, and spatial properties similar to genuine EEG. There have been a few recent efforts to use GANs for EEG data. Recent developments in CNN, as well as successful deployments for EEG tasks, prompted us to select CNN over frequently used autoregressive models for time-series data creation. CNN-based GANs are composed of two NN: the generator (G) and the discriminator (D). 

\begin{figure}[htbp]
    \centering
    \includegraphics[width=7in]{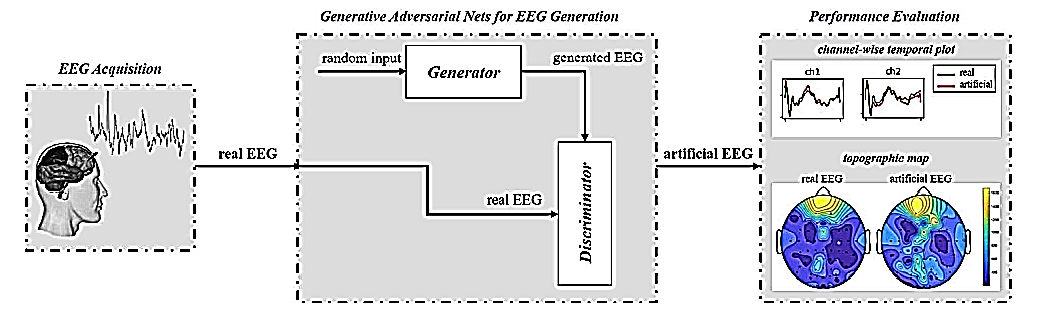}
    \caption{The overall framework of EEG generation using GANs\cite{fahimi2019towards}.}
    \label{gan}
\end{figure}

The overall concept is influenced by game theory, in which two players strive to beat one other.
G's goal in GANs is to produce synthetic data, whereas D's task is to determine which samples are actual and which are generated. G's training goal is to ultimately deceive the discriminator so that the discriminator can no longer differentiate between the actual and produced data. In other words, the produced samples are quite similar to the actual samples. The D and G neural network is trained using actual EEG data. The trained G network is then fed random noise as input. 

As a consequence, the G network creates the simulated EEG from random input. It is therefore critical to assess the performance of GANs by extracting and evaluating the properties of the produced EEG. To do so, the actual and produced EEG of each channel were plotted over time.
In addition, the topographic maps of the real and produced EEGs were computed using the fast furrier transform (FFT) to improve interpretation of the data. Figure \ref{gan} depicts an overview of the suggested technique.

\subsection{Autoencoders}
The term "autoencoding" refers to a data compression technique in which the compression and decompression functions are 1) data-specific, 2) lossy, and 3) learnt automatically from examples rather than created by a person. Furthermore, neural networks are utilised to implement the compression and decompression functions in virtually all instances where the word "autoencoder" is used. 
\begin{figure}[htbp]
    \centering
    \includegraphics[width=7in]{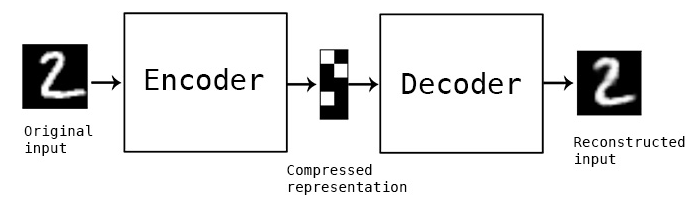}
    \caption{Frame work for Autoencoders}
    \label{frame}
\end{figure}
Autoencoders are a form of Feedforward Network in which the input and output is similar. Autoencoders compress the data into a lower-dimensional form, use to recreate the output. The code, also known as the latent-space representation, is a compact “summary” or “compression” of the input. Encoder, code, and decoder are the three parts of an autoencoder. The encoder compresses the input and generates the code, which the decoder subsequently uses to recreate the input. Three things are needed to create an autoencoder: an encoding technique, a decoding technique, and a loss function which compares the output to the target. Mostly used loss function is mean square error(MSE). Figure \ref{frame} shows the framework for the autoencoders. 
\bgroup
\def\arraystretch{1.5}
\begin{table}[htbp]
\caption{Implementation of autoencoder on the CNN model }\label{auto}
\begin{center}
\begin{tabular}{l l l}
\hline
\hline
\textbf{Layer (type)} &\textbf{Output Shape} &\textbf{Param \#}\\
\hline
\hline
InputLayer &[(None, 24,24,1)] &0\\
\hline
Conv2D &(None,24,24,16) &160\\
\hline
MaxPoolingLayer &(None,12,12,16) &0\\
\hline
Conv2D &(None,12,12,8) &1160\\
\hline
MaxPoolingLayer &(None,6,6,8) &0\\
\hline
Conv2D &(None,6,6,8) &584\\
\hline
MaxPoolingLayer &(None,3,3,8) &0\\
\hline
Conv2D &(None,3,3,8) &584\\
\hline
UpSamplingLayer &(None,6,6,8) &0\\
\hline
Conv2D &(None, 6,6,8) &584\\
\hline
UpSamplingLayer &(None,12,12,8) &0\\
\hline
UpSamplingLayer &(None,24,24,8) &0\\
\hline
Conv2D &(None, 24,24,1) &73\\
\hline
\hline
\end{tabular}
\end{center}
\label{table1}
\end{table}
\egroup

Autoencoders were implemented on the CNN model, as illustrated in Table \ref{auto}. In this case, an input of 24x24 is passed to the encoder, which compresses the picture to generate an image with a compressed dimension of (3x3), which is then passed to the decoder to duplicate the input.
Before training an autoencoder, four hyperparameters were explored:
\begin{itemize}
    \item \textbf{Code size:} number of nodes in the middle layer. \item \textbf{Number of layers:} The autoencoder can have as many layers.
    \item \textbf{Number of nodes per layer:} Since the layers are piled one after the other, this autoencoder design is known as a stacked autoencoder. Stacked autoencoders are typically shaped like a "sandwich." With each consecutive layer of the encoder, the number of nodes per layer drops and then rises in the decoder.
    \item \textbf{Loss function:} MSE was used for multi-class classification  and for binary class, Binary cross-entropy was used as loss function.
\end{itemize}

\section{Proposed Methodology}
The section discusses about data set, followed by the methodology for electrode selection. Further, the pre- processing approaches have been elaborated, followed by artifact identification and rejection. Finally, the feature extraction methodology in several domains has been deliberated in the further section.
\subsection{Dataset}
The paper focuses on upper arm amputees and there are different dataset available\footnote{http://bnci-horizon-2020.eu/database/data-sets} \footnote{http://www.bbci.de/competition/iv/}.
The data set utilized has been discussed below\footnote{https://www.physionet.org/content/eegmmidb/1.0.0/}. 

\subsubsection{EEG Motor Movement/Imagery Dataset}
\begin{figure}[htbp]
\centering
\includegraphics[width=5in]{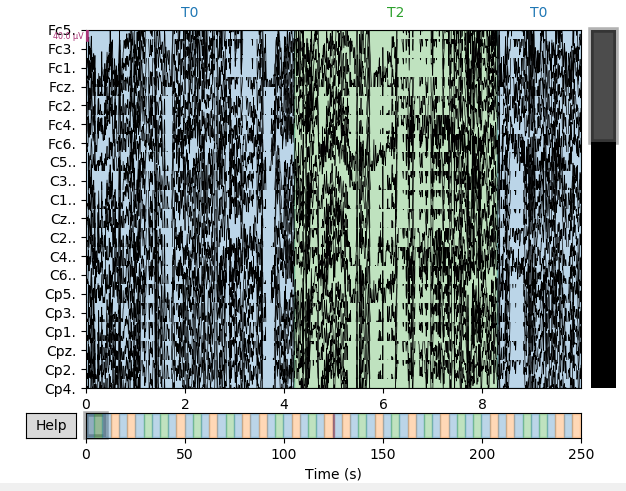}
\caption{ Raw EEG signals for motor imagery class for left and right hand movements}
\label{imag}
\end{figure}

\begin{figure}[htbp]
    \centering
    \includegraphics[width=5in]{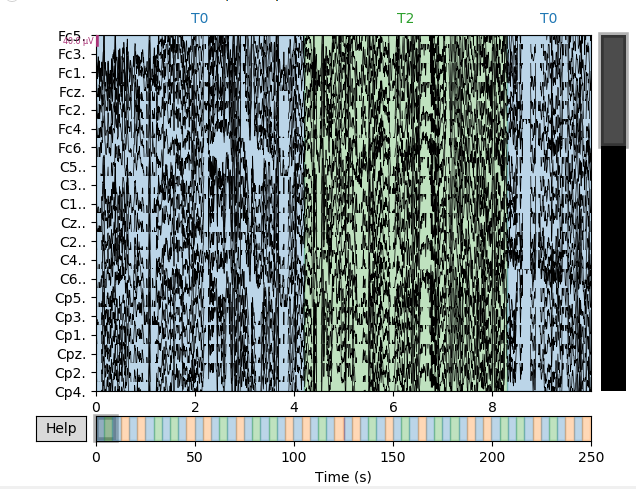}
\caption{Raw EEG signals for motor Execution class for left and right hand movements}
\label{exec}
\end{figure}

\begin{figure}[htbp]
\centering
\includegraphics[width=6.4in]{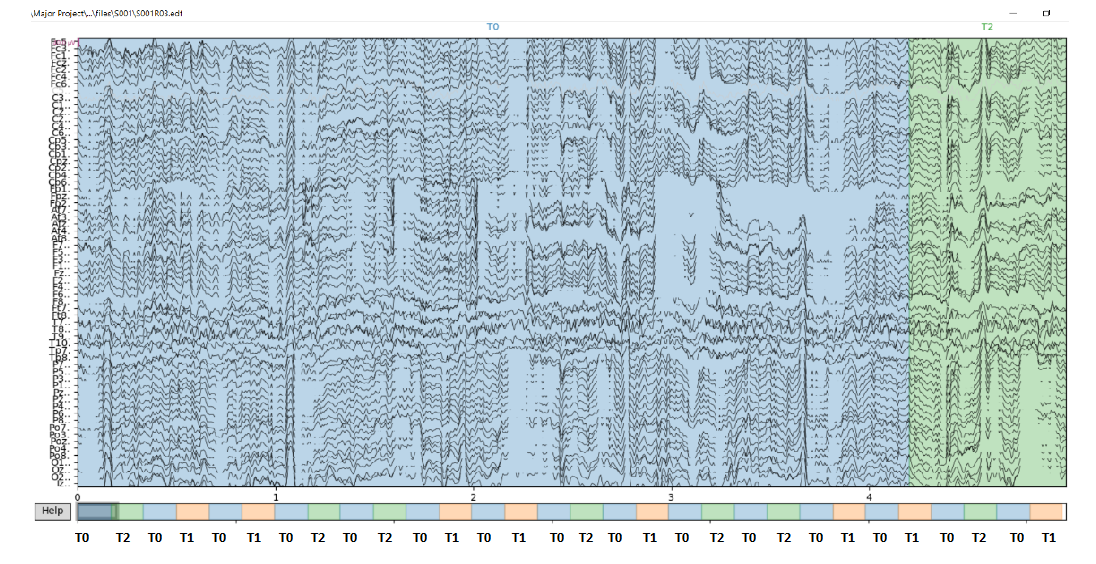}
\caption{Image represents the annotations T0, T1, T2}
\label{both}
\end{figure}

In this paper, an EEG Motor Movement/Imagery dataset was employed. The samples were gathered using a BCI 2000 device on 109 participants. Three exercises were performed by the subjects: rest (T0), left hand movement (T1), and right hand movement (T2). Each experiment consisted of 15 repetitions, with T0 being followed by a visual stimulus, with T1 or T2 being chosen at random. The data was in EDF+ format, with 64 EEG channels that were captured at 160 samples per second.
Figure \ref{imag} and Figure \ref{exec} show the raw EEG signals for motor Imagery and motor execution class respectively for both right hand and left hand movements. Figure \ref{both} shows annotations T0, T1, T2. Here T0 depicts that the subject is at rest,while the T1 and T2 period replicates to the onset of real or imagined left first movement or right fist movement respectively~\cite{zhang2020arder}. 
Here each subject was asked to perform 14 tasks. Among 14 tasks only 6 tasked were considered. Here it is to observed that 3,7 and 11 runs represent motor execution class while 4,8 and 12 runs represent motor imagery class.\\


\subsubsection{Montage}
\begin{figure}[htbp]
    \centering
    \includegraphics[width=5.5in]{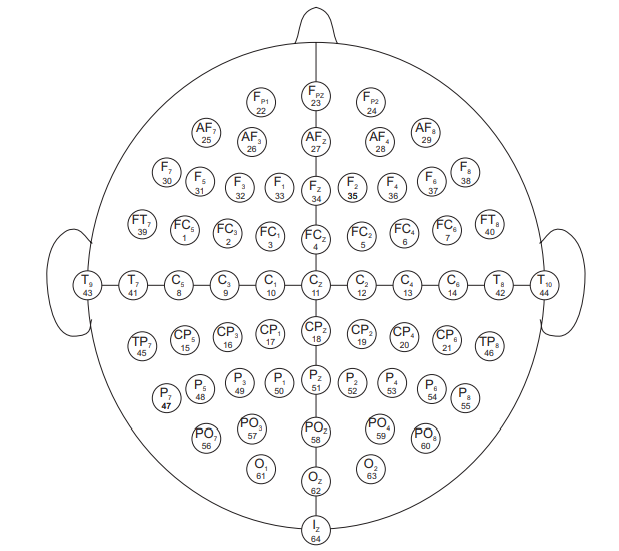}
    \caption{Placement of 64 electrodes in 10-10 international manner}
    \label{placement}
\end{figure}

64 electrodes were used for recording EEG signals using 10-10 international system. 
Figure \ref{placement} depicts the placement of 64 electrodes in a 10-10 international pattern, with the exception of Nz, F9, F10, FT9, FT10, A1, A2, TP9, TP10, P9, and P10. The numbers underneath each electrode name represents the sequence in which they occur in the recordings.  

\subsection{Selection of electrodes}
Proper selection of electrodes is important as it can alter the accuracy~\cite{zhang2019classification}. The paper explained the flow of information through electrodes at certain intervals. Others~\cite{SUN20081663}, on the other hand, they tried selecting different electrodes randomly and obtained different accuracy for randomly selected electrodes. But this method may not give the proper classification.
The subsequent part explains, the flow of information at various points in time, as stated in the study \cite{zhang2019classification}. During the experiment. At \textbf{t=0 sec}, the individuals are exposed to a visual stimulus. Sensor pairs in the anterior-frontal \textbf{(AF3-AF4 and AF7-AF8)}, parietal-occipital \textbf{(PO3-PO4 and PO7-PO8)}, as well as occipital \textbf{(O1-O2)} lobes revealed the greatest traits at this stage. \\

The \textbf{P300 wave} is a form of \textbf{Event-Related Potential(ERP)} that is thought to be prominent in decision-making and typically lasts for 250 ms to 500 ms on the commencement of visual input. Another parameter \textbf{Simple reaction Time (SRT)} represents the time lag between visual stimulus and response. According to the discussion, the estimated time of SRT was $231\pm27ms.$ At \textbf{t=0.25 sec} sensor pairings in the frontal- temporal \textbf{(FT7-FT8)} and temporal-parietal \textbf{(TP7-TP8)} lobes became increasingly prominent. 
The visual information is sent via two distinct streams, namely the ventral and dorsal streams~\cite{de2009handbook}. The ventral stream travels to the temporal cortex, which is primarily responsible for picture identification. The visual stimulus is then used to create the connection between the experiment instructions and the L/R hand movement.
When contrasted to t = 0 sec, sensor pairs in the central-parietal (CP3-CP4 and CP5-CP6) and Parietal lobes (P1-P2, P3-P4, and P5-P6) reveal relevant characteristics at t=0.25 sec. At \textbf{t=0.5 sec}, the electric impulse in the parietal lobes shrinks, indicating the conclusion of information flow in the dorsal stream. As a result, sensor pairs in the central-parietal lobe~\textbf{(CP1-CP2, CP3-CP4, and CP5-CP6)} shows distinct characteristics.
Sensor pairs in temporal \textbf{(T7-T8 and T9-T10)}, frontal \textbf{(F7-F8)}, and anterior-frontal \textbf{(AF7-AF8)} lobes revealed discriminant for L/R hand motions at \textbf{t=0.75 sec.}.The primary cortex is accountable for the execution of all voluntary actions~\cite{ward2019student}. Therefore, electrodes near the motor cortex area was considered. As a result, after understanding the flow of information through each electrode. Table 2 shows the selected electrode for the study \cite{zhang2019classification}.
\bgroup
\def\arraystretch{2}
\begin{table}[htbp]
\label{sample}
\caption{Selection of electrodes}
\begin{center}
\begin{tabular}{c c c c c }
\hline
\hline
\textbf{Sr.No.}  &\textbf{Sensor Pair} &\textbf{Brain Lobe}  &\textbf{Sensor Pair} &\textbf{Brain Lobe}  \\
\hline
\hline
1 &FC5-FC6 &Frontal Central   &F5-F6 &Frontal\\
\hline
2 &FC3-FC4 &Frontal Central   &F7-F8 &Frontal\\
\hline
3 &FC1-FC2 &Frontal Central   &F1-F2 &Frontal\\
\hline
4 &C3-C4 &Central   &T7-T8 &Temporal\\
\hline
5 &CP5-CP6 &Central Parietal   &T7-T8 &Temporal\\
\hline
6 &CP3-CP4 &Central Parietal   &T9-T10 &Temporal\\
\hline
7 &CP1-CP2 &Central Parietal   &TP7-TP8 &Temporal
Parietal\\
\hline
8 &FP1-FP2 &Frontal Parietal   &P5-P6 &Parietal\\
\hline
9 &AF7-AF8  &Anterior Frontal  &P3-P4 &Parietal\\
\hline
10 &AF3-AF4  &Anterior Frontal  &P1-P2 &Parietal\\
\hline
11 &PO7-PO8 &Parietal Occipital  &PO3-PO4 &Parietal Occipital \\
\hline
12 &O1-O2 &Occipital  &Cz-Cpz &central electrodes\\
\hline
\hline
\end{tabular}
\end{center}
\end{table}
\clearpage
\subsection{Pre-processing}
After obtaining the EEG data, these waves are pre-processed to decrease noise in the EEG signals and to increase particular brain information of certain activity patterns from the raw EEG recordings.
After selecting 46 electrodes they were pre-processed using two filters as mentioned below.
MNE package is a dedicated package for EEG signal processing, EEG artifact detection and removal, and source localization, it is been used exclusively for the processing and visualization of EEG signals.
\begin{enumerate}
\item \textbf{Notch filter:} 
It removes the power line interference using a notch filter of 50 Hz. 
\item \textbf{Band-pass filter:} 
It was used to let frequencies ranging from 0.5 to 40 Hz to pass through, decreasing artifacts and denoising. The range chosen also keeps the critical information safe~\cite{benzy2019classification}.
\end{enumerate}

\begin{figure}[htbp]
    \centering
    \includegraphics[width=6in]{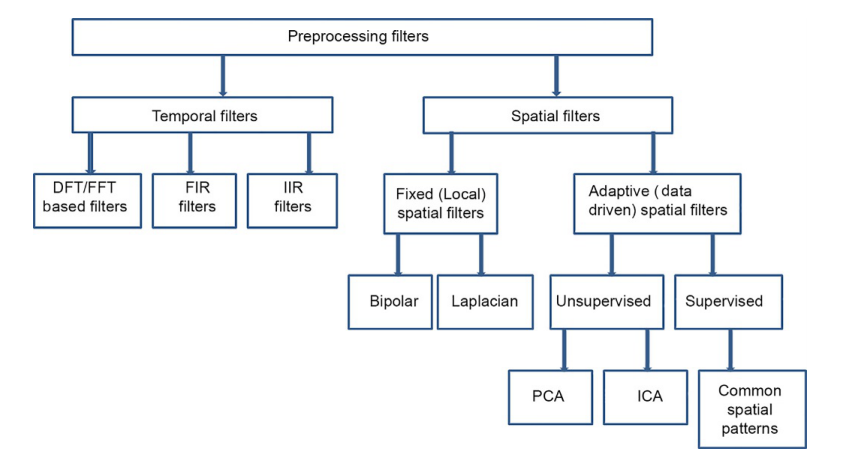}
    \caption{Categorization of pre-processing filters\cite{bansal2019eeg}.}
    \label{preprocessing}
\end{figure}

\subsection{Artifacts Detection and Rejection}
The acquired EEG signals are usually corrupted with different artifacts and it is really important to reject them. 
Removal of artifacts is an extremely critical task. Two techniques were used for the study. First, the signals were inspected using EEGLAB tool. EEGLAB tool \footnote{https://eeglab.org/tutorials/} is an inbuilt tool in MATLAB having functions that helped to understand EEG signals and to determine the artifacts more concretely.
Before applying ICA on the data to be cleaned, it was observed that a high pass filter between 1 to 2 Hz and no low-pass filter appears to be the best choice to enhance ICA disintegration's. ICA is susceptible to low-frequency tones and hence requires the data to be high-passed and separated before fitting. 
The \textbf{FASTICA} technique was used for 46 components and was applied to the data. The number of components in this case refers to the number of main components (from the pre-whitening PCA stage) that are provided to the ICA algorithm during fitting. The ICA components are then sent into the ICA algorithm, which produces the un-mixing matrix.

Figure \ref{ICA_1} and Figure \ref{ICA_2} show the topological plots for all 64 channel using EEGLAB tool which also shows the labels indicating which type of artifact. using visual inspection one can reject contaminated IC's but this method consumes more time. Here, there are 109 subjects and inspecting each and every subject is a complex process and hence it is difficult to use the EEGLAB tool for the elimination of the artifacts.
\begin{figure}[htbp]
    \centering
    \includegraphics[width=7in]{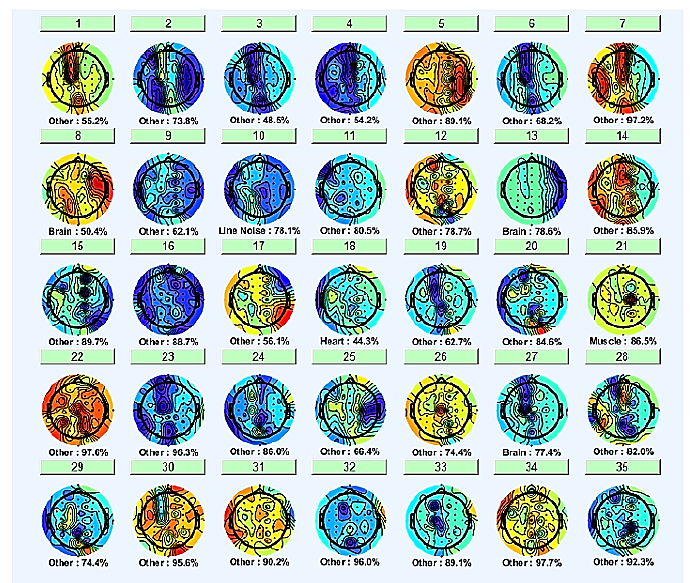}
    \caption{Topological ICA plots for first 35 channels for the first subject \cite{EEGLAB}.}
    \label{ICA_1}
\end{figure}
\begin{figure}[htbp]
    \centering
    \includegraphics[width=7in]{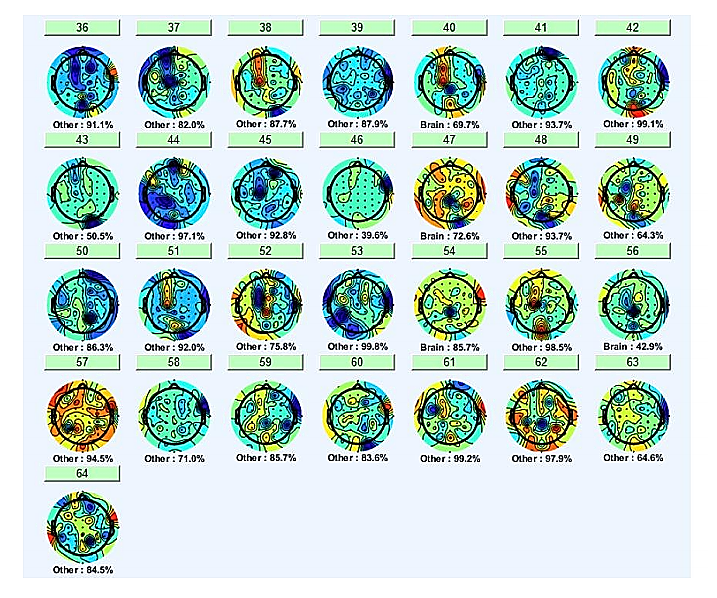}
    \caption{Topological ICA plots for other remaining channels for the first subject \cite{EEGLAB}.}
    \label{ICA_2}
\end{figure}
\clearpage

\subsubsection{ICA in MNE-Python}
The novel algorithm is implemented to remove the undesired signals from the EEG signal and to obtain clean signal. The Python MNE module is used to build the method. The block diagram for ICA-based artifact removal is shown in the Figure \ref{ica_mne}.
\begin{figure}[htbp]
    \centering
\includegraphics[width=6in]{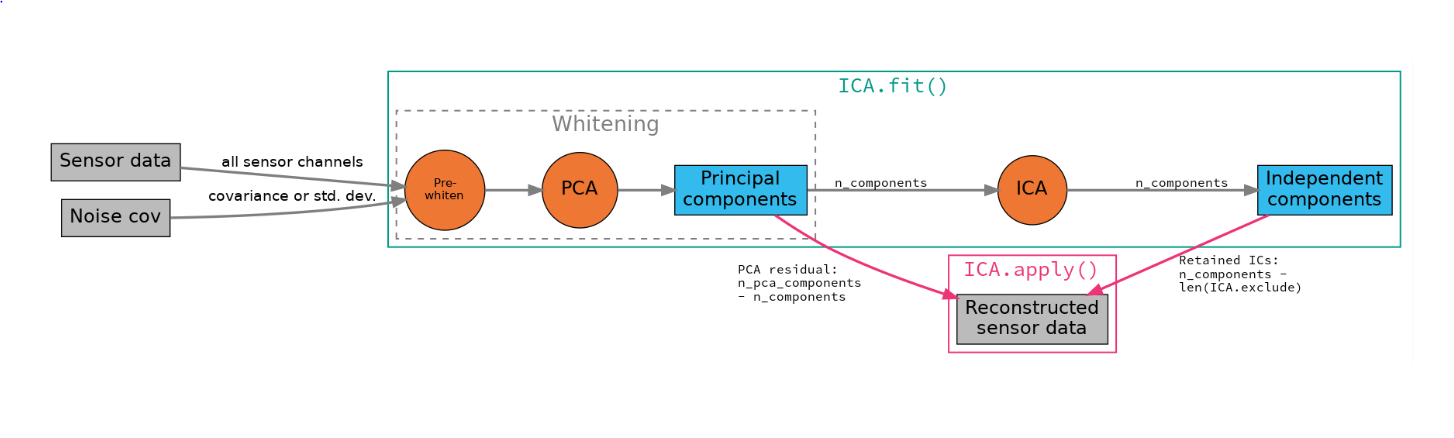}    
    \caption{Block diagram for ICA analysis in MNE-Python\cite{mne_ica}}
    \label{ica_mne}
\end{figure}

Following are steps to detect artifacts and clean EEG signal:
\begin{enumerate}
    \item ICA Fit operation
    \item ICA detect
    \item Applying algorithm on IC's to exclude contaminated IC's
    \item ICA exclude
    \item ICA apply
\end{enumerate}
Figure \ref{icaalgo} demonstrates the methodology for detecting and removing ICA-based artifacts. As it is previously explained, ICA must be routed via a 1 Hz high pass filter. Before using the fit operation PCA components must be defined which will statistically un-mix the signals and obtain various components related to the signals. As discussed earlier, the numerous IC's related to EEG signals were examined using EEGLAB tool in MATLAB. Different independent Components related with EEG signals were examined which were contaminated with various types of signals such as: 1. Power line signal 2. Ocular Signal 3. Muscle related signals and so on.
\begin{figure}[htbp]
    \centering
    \includegraphics[width=7in]{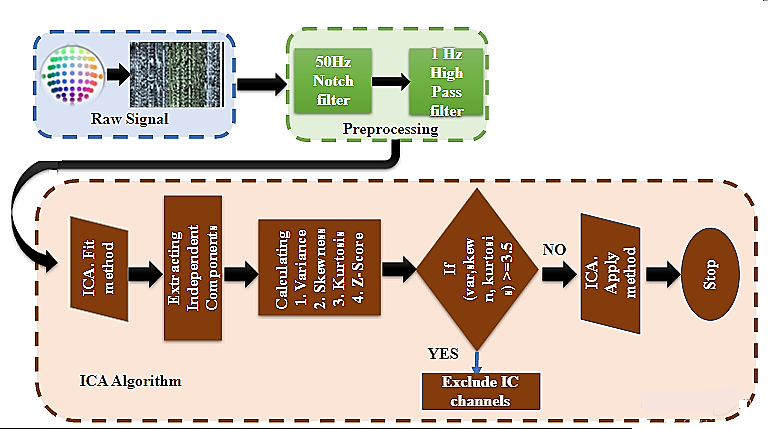}
    \caption{Algorithm for artifact detection and removal}
    \label{icaalgo}
\end{figure}
It is proposed that PCA components must be smaller than the number of channels chosen for the investigation. 46 channels are chosen for the study. Hence Using the ICA.fit approach, it fitted 30 PCA independent components. The Independent components can be seen as illustrated in Figure \ref{diff}. Neurological experts visually examine these waves related to Independent Component and can detect various kind of artifacts linked with the EEG signals. Using the above approach tainted ICs can be ruled out. But the complexity increases as the number of subject increases. As a result, the novel approach was employed which is discussed in the further section. 

\begin{figure}[htbp]
    \centering
\includegraphics[width=5in]{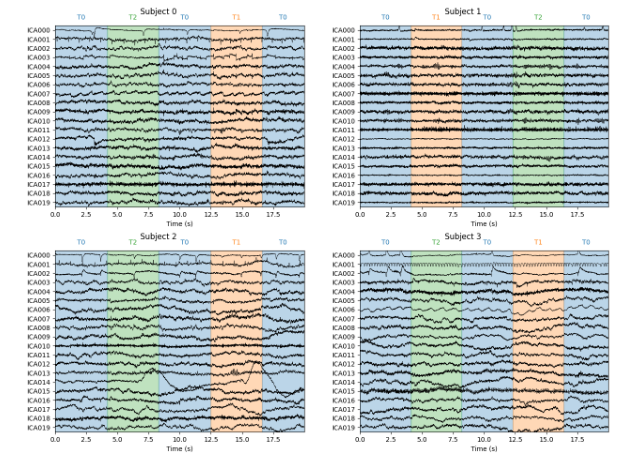}
    \caption{ICA components for subject 0, 1, 2 and 3}
    \label{diff}
\end{figure}

Two algorithms were implemented for the analysis. Firstly, monitored the IC's related to EEG channels and subsequently calculated statistical parameters such as 1. Variance 2. Skewness 3. kurtosis 4. Z-score for kurtosis of IC's. 
The $2^{nd}$ strategy utilized the MNE package's inbuilt command, ICA detect, which has certain attributes such as variance, Kurtosis, Skewness, and Person's linear Correlation coefficient with a specified threshold. The above parameters' thresholds are set in accordance with ~\cite{dammers2008integration}. In article~\cite{dammers2008integration} they have calculated the threshold values for these attributes, which is discussed in more detail in further section.\vspace{8pt}\\
\subsection{Artifact Detection Measures}
\begin{enumerate}
    \item \textbf{Variance}\\
    The large value of Variance intimates the presence of muscular artifacts, eye blinks, or big transient noise signals.
    \item \textbf{Skewness}\\
    It is the third-order central moment that indicates a measure of the degree of asymmetry of a distribution. Higher value intimates IC's polluted with ocular artifacts.
    \item \textbf{Kurtosis}\\
    Peaks in the Waves are characterized by kurtosis value. The greater the value more is the transient muscle or cardiac activity.
    \item \textbf{Entropy}\\
    It denotes the likelihood of identifying amplitude values in a magnitude distribution in $i^{th}$ IC. Higher entropy values symbolize more complications.
    \item \textbf{Pearson's Linear Correlation}\\
    To discriminate between ocular and cardiac activity in groups of ICs, Pearson's Linear was used as Correlation coefficients between each IC. 
\end{enumerate}
Threshold values for the above central moments and linear correlation is $\pm 3.5$  and $>= 0.5$ respectively. 

Variance, kurtosis, and skewness were used as evaluation criteria in the first method. As it was already stated in upper section that kurtosis is a vital parameter. In this section, a brief analysis of kurtosis and its z-score is performed according to \cite{dammers2008integration}. If the value of Kurtosis is greater than 8.83 and z-score of kurtosis greater that 0.23 then IC's suffer from EoG or ocular artifacts.

The second technique employed all the previously described attributes related to ica.detect\_artifacts methods such as eog\_criterion, skewness, kurtosis, variance, and entropy. After going through the documentation it is to be observed that the method ica.detect\_artifact is just for experimentation and further research is required. The in-built technique for identifying faulty components is shown in Figure \ref{ica-det}. This method specifies which channels should be discarded.
Figure \ref{ocular} displays the detection of an eye blink artifacts using in-buit functions~\cite{dammers2008integration}.
\begin{figure}[htbp]
    \centering
    \includegraphics[scale=0.6]{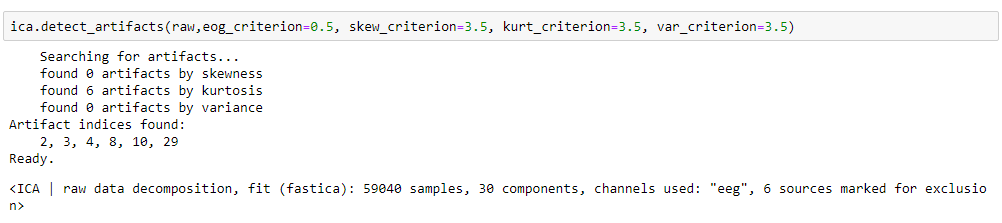}
    \caption{Syntax for ICA detection}
    \label{ica-det}
\end{figure}

\begin{figure}[htbp]
    \centering
    \includegraphics[width=7in]{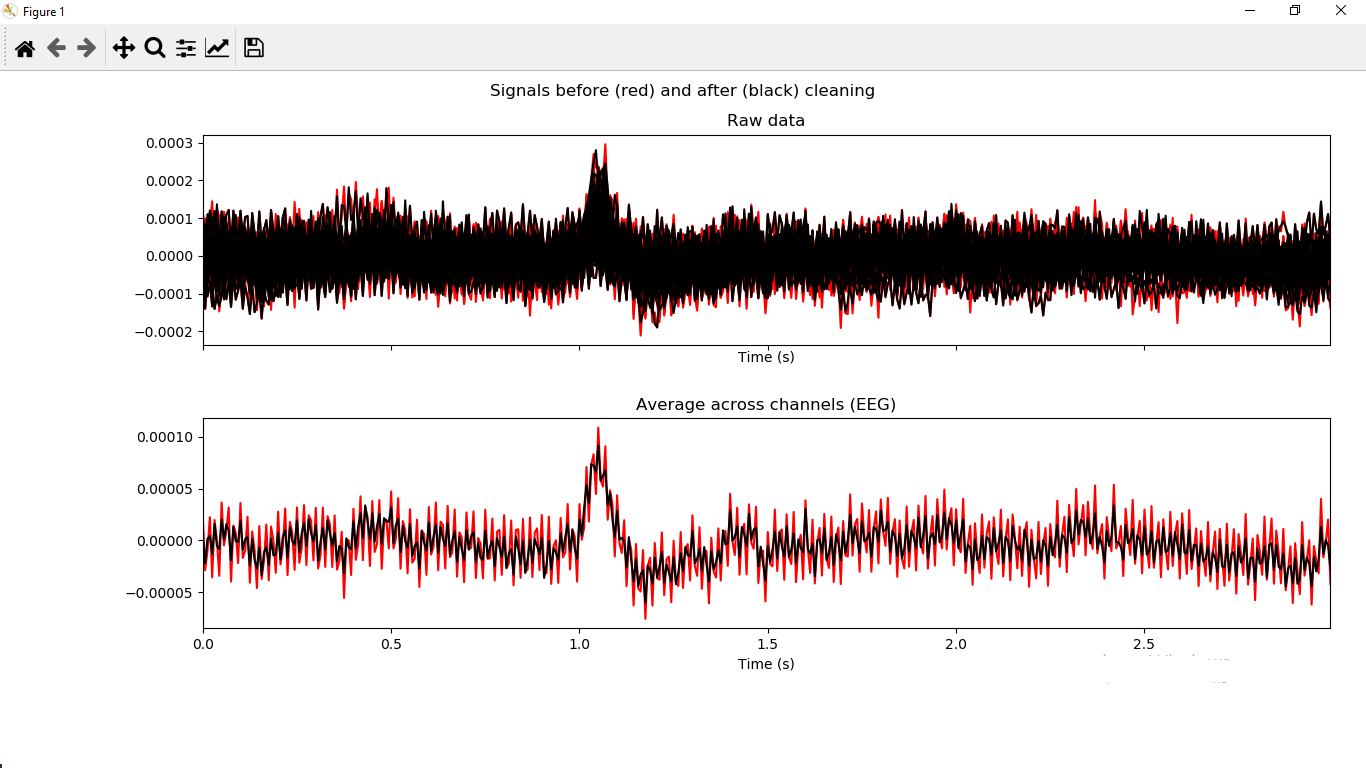}
    \caption{Detection of Ocular artifact}
    \label{ocular}
\end{figure}
\subsection{Signal Segmentation}
In order to develop an acceptable data segmentation method, two major components must be carefully considered: window size and the degree of overlap among two successive window fragments. The window size is a balance between accuracy of classifier and classification reaction time. Classification performance improves as window size grows. Nonetheless, classification size, must be kept as short as possible in order to meet rigorous real-time performance requirements. After examining various window size and segment overlap a decision was made to consider 560 samples per window having a stride of 1 for feature extraction.
This window size reduced the time complexity and improved the outcomes. According to \cite{zhang2019classification}, 2 sec window period and 50\% overlap is considered. However, for the assessment, a window size of 3.5 seconds and the rate at which the window moved was selected as shown in Figure \ref{seg} .

\begin{figure}[htbp]
    \centering
    \includegraphics[width=5.5in]{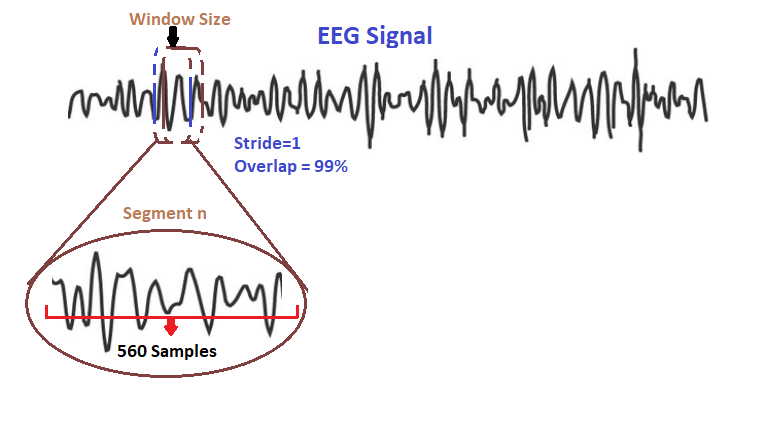}
    \caption{Feature extraction method based on overlapping window.}
    \label{seg}
\end{figure}
\subsection{Feature Extraction}
After detailed discussion on artifacts detection and removal in this section feature extraction techniques were carried. 
The technique of representing obtained and pre-processed EEG data using compressed and linked signal values is known as feature extraction. These features are nothing but feature vectors. EEG signals are non-linear, non-Gaussian, and non-stationary.
Hence the features are obtained from 4 domains of EEG signal representation.   
\begin{enumerate}
    \item \textbf{Time domain depiction of an EEG signal}
    \item \textbf{Frequency domain depiction of an EEG signal}
    \item \textbf{ Time and frequency domain depiction of an EEG signal}
    \item \textbf{Spatial domain depiction of an EEG Signal\cite{de2009handbook}}.
\end{enumerate}
Different State-of-art techniques for feature extraction were observed and conclusion was made to utilize only time and frequency domain. 

\subsubsection{Time domain feature Extraction}
In papers \cite{hossain2020detection}, \cite{zhang2019classification} they have worked on certain statistical features \cite{kaur2017detection}. Five features in the time domain is represented below:
\begin{enumerate}
    \item \textbf{Mean}
            \begin{equation} \label{eq1}
        \begin{split}
        \mu & = \frac{1}{N}\sum_ {i=1}  ^{N} x_{i}
        \end{split}
        \end{equation}
        Solving equation (\ref{eq1}) for mean in a moving average manner having window size of 560 samples in one window.
    \item \textbf{Variance}
        \begin{equation} \label{eq2}
        \begin{split}
        \alpha ^{2}=\dfrac{1}{N}\sum ^{N}_{i=1}\left( x_{i}-\mu \right) ^{2}
        \end{split}
    \end{equation}
     Solving equation (\ref{eq2}) for variance in a moving window having window size of 560 samples in one window.
    \item \textbf{Skewness}
    \begin{equation} \label{eq3}
        \begin{split}
        S & = \begin{aligned}\dfrac{1}{N}\sum ^{N}_{i=1}\left( x_{i}-\mu \right) ^{3}\\ \overline{\left( \dfrac{1}{N-1}\sum ^{N}_{i=1}\left( x_{i}-\mu \right) ^{2}\right) }^{3/2}\end{aligned}
        \end{split}
    \end{equation}
    Equation (\ref{eq3}) is the third-order central moment that indicates a measure of the degree of asymmetry of a distribution called Skewness~\cite{7319296}. 
    
    \item \textbf{Kurtosis}
    \begin{equation} \label{eq4}
        \begin{split}
        K & = \begin{aligned}\dfrac{1}{N}\sum ^{N}_{i=1}\left( x_{i}-\mu \right) ^{4}\\ \overline{\left( \dfrac{1}{N-1}\sum ^{N}_{i=1}\left( x_{i}-\mu \right) ^{2}\right) }^{2}\end{aligned} - 3
        \end{split}
    \end{equation}
    Peaks in the Waves are characterized by kurtosis value. Here kurtosis is a proven statistical feature for classification. Equation (\ref{eq4}) shows the formula for Kurtosis. 
    \item \textbf{Absolute Area under signal}
    \begin{equation} \label{eq5}
        \begin{split}
            A=\int ^{b}_{a}\left| f\left( x\right) \right| dx
        \end{split}
    \end{equation}
   Equation (\ref{eq5}) shows the formula for area under signal which is one of the vital feature for classification.
\end{enumerate}

\clearpage
\subsubsection{Frequency domain feature Extraction}
Certain specific frequency domain characteristics were chosen from a review paper~\cite{boonyakitanont2020review}.
Frequency domain analysis is also necessary since a frequency representation of an EEG signal gives some helpful information about the signal's patterns. The \textbf{Power Spectral Density (PSD)} equation (\ref{eq6}) and the normalized PSD (by the total power) are commonly used to derive characteristics encapsulating the power partition at each frequency.
This section summarises the PSD characteristics for the extraction purpose.

\begin{equation}
    \begin{split}\label{eq6}
        S_{xn}\left( W\right) =\lim _{T\rightarrow \infty }E\left[ \left| \widehat{X}\left( W\right) \right| ^{2}\right]
    \end{split}
\end{equation}

\begin{enumerate}
    \item \textbf{Peak Frequency}\\
    It is the frequency at which the PSD of the highest average power in the full-width-half-max (FWHM) band has the greatest magnitude.
    
    \item \textbf{Peak Amplitude}\\
    Here Peak Amplitude represents the highest value of spectral density.\\
\end{enumerate}

Figure \ref{psd_ext} and Figure \ref{psd_img} show the Power spectral Density for both motor Imagery and Execution. Tabel \ref{table1} shows the values associated with the feature vectors in raw form these are not yet normalized. 

\subsubsection{Feature Selection approach}     

\begin{figure}[htbp]
    \centering
    \includegraphics[width=3.5in]{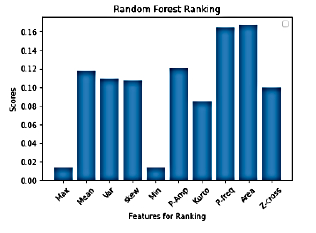}
    \caption{Feature Ranking using Random Forest}
    \label{lab_1}
\end{figure}
 To investigate the involvement of several features, The paper \cite{zhang2019classification} estimated the significance of each feature through the use of Random Forest (RF) for feature ranking. And according it, the first top three features were skewness, mean, and area under signal~\cite{boonyakitanont2020review}. Figure \ref{lab_1} displays the relevance of each characteristic when Random Forest (RF) is used to rank features. Here it is been examined that, the features like mean, skewness, variance, kurtosis, area under curve, peak frequency, and peak amplitude have a role in enhancing the performance of a classification model. It can be seen that characteristics like min, max, and zero crossing have less relevance when compared to the selected parameters. As a result, it can observed that the most vital key features were chosen for classification.
\begin{figure}[htbp]
    \centering
    \includegraphics[width=5in]{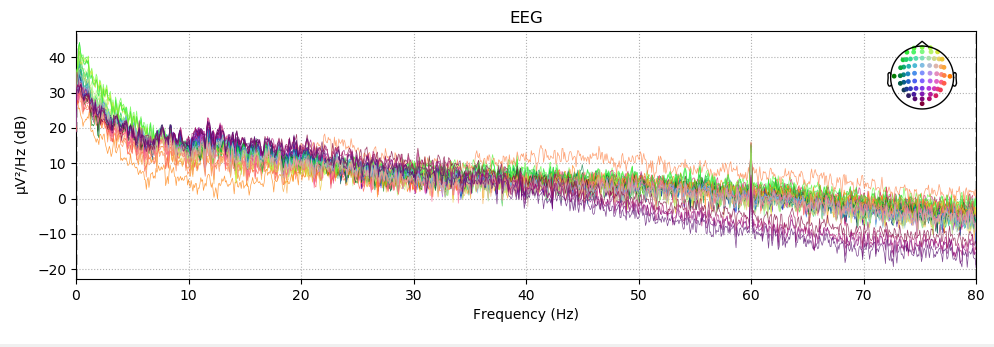}
    \caption{Power Spectral Density for Motor execution class}
    \label{psd_ext}
\end{figure}

\begin{figure}[htbp]
    \centering
    \includegraphics[width=5in]{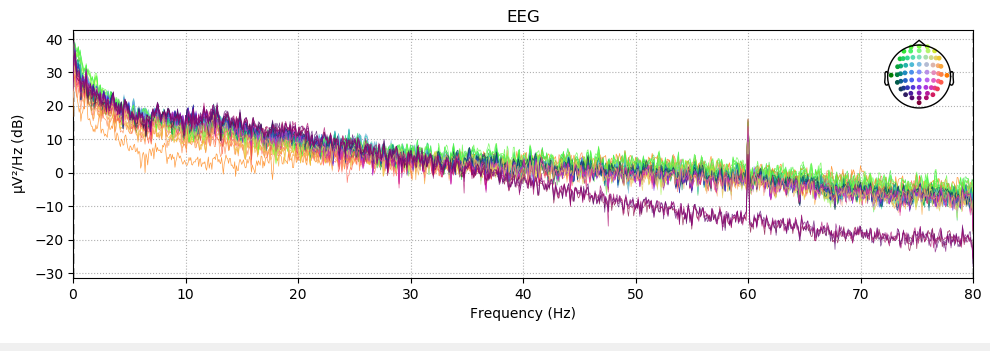}
    \caption{Power Spectral Density for Motor Imagery Class}
    \label{psd_img}
\end{figure}

\subsection{Normalization}
The normalization with Z-score is represented by Equation (\ref{eq7})
\begin{equation}\label{eq7}
    \begin{split}
        z=\dfrac{x-\mu }{\sigma}
    \end{split}
\end{equation}
 The features were normalized using Z-score for this an in-built library Sklearn was used. Sklearn has a class called \textbf{StandardScalar} which contains fit\_transform to Normalize the values. 
\begin{itemize}
    \item \textbf{z= Standard value}
    \item \textbf{x= observed value}
    \item \textbf{$\mu$= mean of the sample}
    \item \textbf{$\sigma$= standard deviation of the sample}
\end{itemize}

\subsection{Proposed Model}
Here the proposed model used Multi-layer Perceptron Artificial Neural Networks (MLP) \cite{jain2019iglu}.
The Neural Network model used was basically a sequential model having:
\begin{itemize}
    \item A input layer having 700 Perceptrons having \textbf{RELU} as activation function.
    \item Input size depends upon the number of features utilized, Seven features in the data-set were utilized, five in time domain and two from frequency domain. Hence the input size=(input\_shape,)
    \item Six hidden layers were implemented having 128, 128, 64, 64, 32, 32 neurons having relu as activation function for each. By default kernel initializer is \textbf{'glorot uniform'} and bias initializer is  zeros. One can change according to its algorithm.
    \item Number of units(neurons) in the output layer depends upon the number of classes. The classifier used was \textbf{softmax}. Four classes are discussed below: 
    \begin{enumerate}
        \item \textbf{Left \& Right hand movement for motor execution}
        \item \textbf{Left \& Right hand movement for motor imagery}
    \end{enumerate}
    \item Adam as optimizer was utilized having $lr=0.001,beta_1=0.9, beta_2=0.999,epsilon=1e-07.$ Sparse Categorical Cross-entropy was used as a loss function. 
    \item It implemented two metric parameters one \textbf{Accuracy} and other \textbf{Mean Square Error} for evaluating the model. 
    \item From Figure \ref{model} it can be referred, the architecture of proposed ANN model. 
\end{itemize}
\begin{figure}[htbp]
    \centering
    \includegraphics[width=5in]{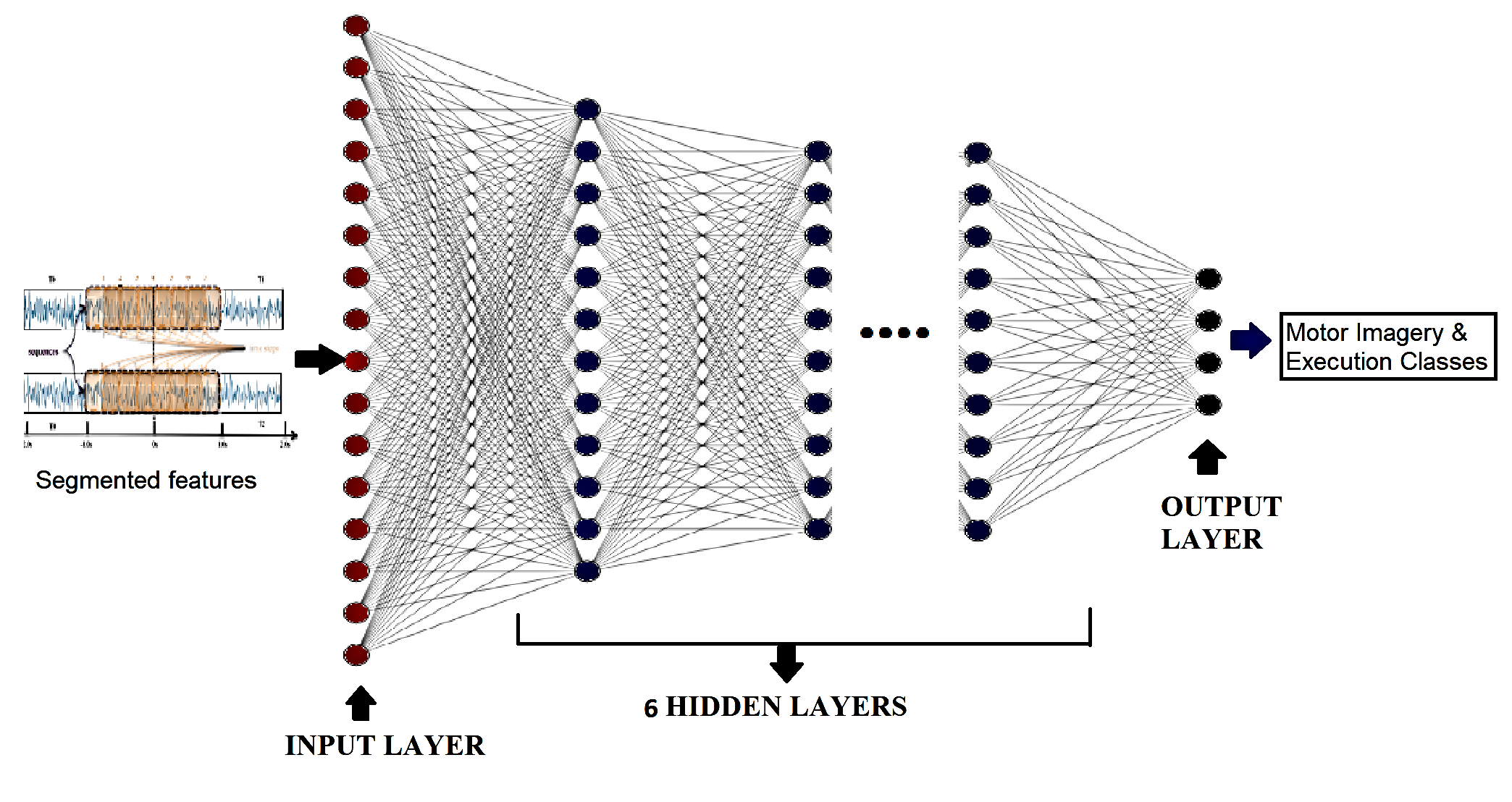}
    \caption{Proposed ANN-model architecture}
    \label{model}
\end{figure}

\subsection{Deep Learning}
In this section, the implementation of deep learning was observed on previously discussed data-set.
In referred paper\cite{li2020electroencephalography} to enhance the domain knowledge and to implement it simultaneously. 
Deep learning algorithms have been attempted since they provide better performance due to their automatically learning capability through features from large size of data~\cite{pancholi2021robust}.

The sliding window technique with a distinct window size of 21 having a stride of 21 hence converting into small fragments. Combining these small fragments created an 2D-Matrix. The deep learning algorithm was implemented only on first subject.
First the pre-processing and normalization was done on the data-set and converted it into an image.
\subsubsection{Model Architecture}
\begin{itemize}
    \item Using the segmentation approach, transformed the 1D signal into a 2D format. The image's dimensions were (21, 21, 1).
    \item The image was feed to Batches of Convolution layers having kernel size of (3x3) and also used MaxPooling layers to reduce the dimensional of an image.
    \item Similarly the two more layers arranged in the above mentioned manner having 64 and 128  units. All three CNN layers are converged into one layer called flatten.
    \item After flattening the layers they are interconnected to Fully connected layers called Dense layer having 1024 units activation function as 'relu' and dropout rate as 0.5.
    \item These fully connected layers connected back again to Dense layers of 1024 units and dropout rate as 0.5 followed by output layer of 4 neurons having softmax as classifier. Here categorical cross-entropy and Adam optimizer was implemented and metric parameter has accuracy. 
\end{itemize}

\section{Results and Discussion}
In this part, the performance of the proposed algorithm is discussed on the selected data set. Examination of the results for artifact removal approach was carried further. Later, Investigation of the performance of feature vectors in several domains was carried and choose the optimal domain for testing our algorithm.
The same strategy for 9 participants after successfully removing extraneous signals from EEG recordings and choosing appropriate feature vectors. A brief discussion regarding the outcomes for these subjects are carried out. Lastly, the performance of the proposed model was compared to deep learning model. 

\subsection{Artifact removal using ICA }
 \begin{itemize}
     \item As mentioned in previous sections, for Artifacts Detection, the utilized approach that computed the kurtosis, skewness, variance, and z-score of kurtosis.
 
 \item Table \ref{table3}, \ref{table4} shows the results which displays the contaminated Independent Components for Motor Imagery and Motor Execution respectively. 
 
 \item Threshold for Varaiance, Skewness and Kurtosis is $\pm 3.05$ and Threshold value of Z-score of Kurtosis is 0.23.
 
 \item The contaminated and non-contaminant ICs are shown in two tables below. As an example: It is evident from the independent components in Table \ref{table3} that the ICs 12,13,14,15,18,19 may be excluded. The kurtosis value exceeds the onset value, and the Z-score value exceeds the limit.
 
\item Furthermore, it can be observed that the equivalent results in the case of Motor Execution Class. For Variance of IC's for 18 and 19 is -3.27878 and -3.51884. The values are greater than set threshold. 

\item From the above mentioned approach, bad Independent Components were removed. After discarding the contaminated IC's. An ICA.apply method was performed which mixes all the Independent Components of EEG signal. Hence, it successfully removed artifacts associated with brain wave and obtained clean EEG patterns.

 \end{itemize}
\bgroup
\def\arraystretch{1}
\begin{table}[htbp]
\caption{ICA detection for Motor Imagery }
\begin{center}\label{table3}
\begin{tabular}{c c c c c c}
\hline
\hline
\textbf{Sr.No.} &\textbf{Variance}	&\textbf{Skewness}	&\textbf{Kurtosis}	&\textbf{Z-Score} &\textbf{Result} \\
\hline
\hline
1	&0.034742	&0.108255	&-0.383753	&-0.725933 &*\\
\hline
2	&0.0343778	&0.0130993	&-0.626377	&-0.807453 &*\\
\hline
3	&0.0226487	&0.200304	&-0.775486	&-0.857553 &*\\
\hline
4	&0.0176305	&0.243581	&-0.812049	&-0.869838 &*\\
\hline
5	&0.02656	&-0.450154	&1.34709	&-0.144385 &*\\
\hline
6	&0.0293302	&0.247553	&1.05753	&-0.241674 &*\\
\hline
7	&0.0321314	&0.408744	&0.822696
&-0.320576 &*\\
\hline
8	&0.032728	&0.422592	&0.482036	&-0.435035 &*\\
\hline
9	&0.0287436	&0.463421	&-0.0855045	&-0.625724 &*\\
\hline
10	&0.0253485	&0.400195	&-0.436871	&-0.743781 &*\\
\hline
11	&0.0194755	&0.11406	&-0.856049	&-0.884621 &*\\
\hline
12	&0.0749796	&-2.52961	&8.80182	&2.360347 &\#\\
\hline
13	&0.0543406	&-2.60722	&9.06572	&2.449017 &\#\\
\hline
14	&0.0601778	&-3.06529	&11.5707	&3.290666 &\#\\
\hline
15	&0.0701443	&-1.39425	&4.1624	&0.801538 &\#\\
\hline
16	&0.0664136	&-1.33049	&3.78757	&0.675599 &\#\\
\hline
17	&0.0375335	&-0.707176	&0.721974	&-0.354418 &*\\
\hline
18	&0.0364201	&-1.67454	&3.93	&0.723453
&\#\\
\hline
19	&0.0495347	&-3.35156	&12.4	&3.569295
&\#\\
\hline
20	&0.0473867	&-0.12192	&-0.146448	&-0.646201 &*\\
\hline
\hline
\multicolumn{6}{c}{$*$ Not Contaminated} \\
\multicolumn{6}{c}{\# Contaminated}\\  \hline
\hline
\end{tabular}
\end{center}
\end{table}

\bgroup
\def\arraystretch{1}
\begin{table}[htbp]
\caption{ICA detection for Motor Execution }
\begin{center}\label{table4}
\begin{tabular}{c c c c c c}
\hline
\hline
\textbf{Sr.No} &\textbf{Variance}	&\textbf{Skewness}	&\textbf{Kurtosis}	&\textbf{Z-Score} &\textbf{Result} \\
\hline
\hline
1	&0.0257566	&-1.25491	&1.60076	&-0.304928 &*\\
\hline
2	&0.0256388	&-0.823197	&0.0381866
&-0.698148 &*\\
\hline
3	&0.0185465	&-0.572666	&0.801421	&-0.506080 &*\\
\hline
4	&0.0149535	&-0.405297	&0.745096	&-0.520255 &*\\
\hline
5	&0.0209963	&0.417453	&-0.144257	&-0.744060 &*\\
\hline
6	&0.0227839	&0.00712086	&0.210892	&-0.654687 &*\\
\hline
7	&0.0236176	&-0.458646	&0.588885	&-0.559565 &*\\
\hline
8	&0.0240898	&-0.776216	&0.946091	&-0.469674 &*\\
\hline
9	&0.0212467	&-0.627045	&0.66175	&-0.541229 &*\\
\hline
10	&0.0200463	&-0.275208	&0.331227	&-0.624404 &*\\
\hline
11	&0.016156	&0.00076477	&0.0478366	&-0.695720 &*\\
\hline
12	&0.124933	&-2.91501	&12.3423	&2.398167 &\#\\
\hline
13	&0.0926662	&-3.31493	&14.1188	&2.845225 &\#\\
\hline
14	&0.110129	&-3.31426	&13.7095	&2.742223 &\#\\
\hline
15	&0.098688	&-2.33557	&9.37312	&1.650985 &\#\\
\hline
16	&0.0933006	&-2.35998	&9.61372	&1.711530 &\#\\
\hline
17	&0.044788	&-1.9429	&7.49195	&1.177589 &*\\
\hline
18	&0.0537186	&-3.27878	&13.4315	&0.672277 &\#\\
\hline
19	&0.094339	&-3.51884	&15.0059	&3.068464 &\#\\
\hline
20	&0.0461517	&0.114449	&4.09738	&0.323346 &\#\\
\hline
\hline
\multicolumn{6}{c}{* Not Contaminated} \\
\multicolumn{6}{c}{\# Contaminated}\\ \hline
\hline
\end{tabular}
\end{center}
\end{table}

\subsection{Effect of window size and segment overlap}
The effect of various window sizes and segmentation overlap is shown in Table \ref{effect}. Several segment sizes (0.5, 1.0, 1.5, 2.0, 2.5, 3.5, 4 seconds) and overlap (75 percent and 90 percent) were tested to find the best window size and best overlap for feature extraction. These were used to evaluate the model's performance. Table \ref{effect} provides the best results for a window size of 3.5 seconds that is resulting roughly 560 samples per window size and a stride of 1. Using Table \ref{effect} concluded that selected window size and overlap increased the performance and therefore, these segment parameters were utilized to extract feature vectors.
Google colab was used for testing window size and overlap for feature extraction. All of this testing was done in an online runtime environment. The specifications of the environment was 12GB of RAM and 64GB of storage space. Finally using this approach concluded the optimal segmentation size and fragment overlap.
\bgroup
\def\arraystretch{0.8}
\begin{table}[htbp]
\caption{Effect of window size and overlapping over performance }\label{effect}
\begin{center}
\begin{tabular}{c c c c c c}
\hline
Sr.No. &Window Size &Overlap &Duration for &Accuracy &Validation Accuracy\\
&(sec) &\% &extracting samples &\% &\%\\

\hline
1 &0.5 &90 &3m 52sec &55.39 &31.99\\
\hline
2 &1 &75 &45sec &69.14 &36.56\\
\hline
3 &1 &90 &1m 25sec &55.27 &36.20\\ 
\hline
4 &2 &75 &17.38sec &86.10 &40.65\\
\hline
5 &2 &90 &38.73sec &77.60 &43.49\\
\hline
6 &2.5 &75 &24.76sec &78.25 &37.96\\
\hline
7 &2.5 &90 &1m 2sec &66.55 &39.88\\
\hline
8 &3.5 &50 &3.6sec &90.53 &47.82\\
\hline
\textbf{9} &\textbf{3.5} &\textbf{99} &\textbf{1m 14sec} &\textbf{97.39} &\textbf{94.23}\\
\hline
10 &4 &90 &38.84sec &75.09 &43.66\\
\hline
\end{tabular}
\end{center}
\label{table1}
\end{table}
\egroup
\subsection{Comparison of accuracy for different domains}
After observing the results for artefact removal in the preceding section. The performance of feature vectors were observed in this section using our proposed method. To begin, it is limited to a single subject to determine the accuracy in different domains. The extraction of features and the selection of relevant vectors is a vital step \cite{tadalagi2021autodep}. As a result, rather than picking a domain at random, Conducted some experiments, observed the results, and then made a decision.

Following factors were considered: the statistical metrics are 1. Mean 2. Variance 3.skewness, 4.kurtosis, and 5.area under the peak. The data-set was divided into 80:20 training and validation sets. This also aided in avoiding over-fitting. It was trained using 100 epochs and a batch size of 64 and attained an accuracy of 72.55 \%. While Table \ref{table2} depicts the validation and test accuracy. 
\bgroup
\def\arraystretch{1}
\begin{table}[htbp]
\caption{Accuracy for different sets of features}
\begin{center}\label{table2}
\begin{tabular}{c c c c}
\hline
\hline
\textbf{Domains} &\textbf{Accuracy} &\textbf{Validation Accuracy}  &\textbf{Test Accuracy}  \\
&\textbf{\%} &\textbf{\%} &\textbf{\%} \\
\hline
\hline
Time domain features &72.55 &70.60 &70.21\\
\hline
Frequency domain features &45.90 &45.97 &46.3851\\
\hline
Time and frequency domain features &96.02 &.94.67 &94.431\\
\hline
\end{tabular}
\end{center}
\end{table}
Frequency domain analysis is particularly important since a frequency representation of an EEG signal gives some relevant information about the signal's patterns. Two features of Power Spectral Density: 1. Maximum frequency 2. Peak Amplitude. Using these parameters, the proposed model was trained and obtained an accuracy of 45.90\% for identical hyper parameters such as epochs with 100 and batch-size 64.\vspace{16pt}\\ Eventually integrated the two domains, frequency domain features and time domain features, and attained an accuracy of 95.97 percent. It is apparent that by merging the two domains,  improved the performance. It is important to remember that these features were normalised and that all pre-processing was completed prior to feature extraction. All above techniques were applied all on 9 participants for validation, and the outcomes are discussed in the following section.

\bgroup
\def\arraystretch{1}
\begin{table}[htbp]
\caption{Performance Metrics for time-frequency domain}
\begin{center}\label{met}
\begin{tabular}{c c c c c}
\hline
\hline
&\textbf{Precision} &\textbf{Recall} &\textbf{F1-Score}  &\textbf{support}  \\
\hline
\hline
Class 0     &0.9594    &0.9186    &0.9386     &17132\\
 \hline
Class 1     &0.9349    &0.9670    &0.9507     &17248\\
\hline
Class 2     &0.9561    &0.9371    &0.9465     &17278\\
\hline
Class 3     &0.9272    &0.9530    &0.9399     &17155\\
\hline
    accuracy        &           &      &0.9440     &68813\\
\hline
\hline
\end{tabular}
\end{center}
\end{table}

 The Accuracy v/s Epochs and Loss v/s Epochs plots for the Time-frequency domain combined result, time domain performance, and frequency domain outcome are shown in the Figure \ref{acctime}. The validation plot is virtually identical to the real plot, indicating that the model is not over-fitting. These graphs also suggested in determining the learning rate of the Adam optimizer, which is one of the hyper parameters. Table \ref{met} shows performance Metrics for Subject one in time and frequency domain. 

\begin{figure}[htbp]
    \centering
    \includegraphics[width=7in]{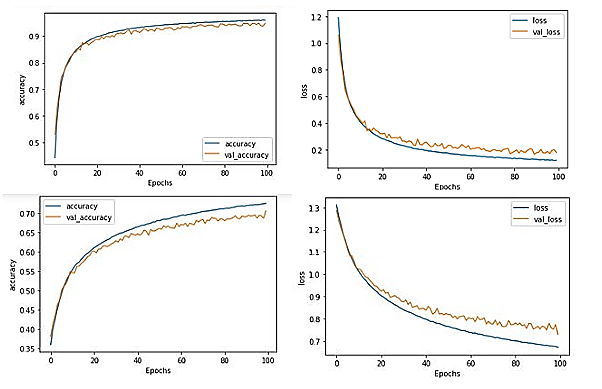}
    \includegraphics[width=7in]{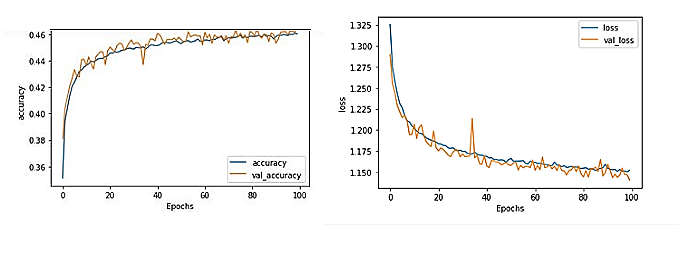}
    \caption{Graph for Accuracy and loss vs epochs for Time-Frequency domain, Time domain and frequency domain}
    \label{acctime}
\end{figure}
\subsection{Performance evaluation on different subjects.}
Till now, a single subject was tested and came to conclusions concerning channel selection, various artefact removal strategies, different domain feature extraction, and the relevance of each on an individual scale for feature. The single subject's results was validated employing the intra-subject validation methodology. Each subject's EEG data went through the same method and in Intra-subject Validation schemes, each individual subject's EEG data will be processed.  Each subject will be evaluated separately. Same methodology was applied on these subjects, Only time and frequency domains were considered, which gave good results earlier. Table \ref{subject} shows performance of different subjects and mean accuracy obtained was 94.77\%. Figure \ref{9sub} depicts the performance of nine subjects in terms of accuracy and loss v/s epochs. Figure \ref{9subv} depicts the performance of nine subjects in terms of Validation Accuracy and loss v/s Epochs. 

\def\arraystretch{0.8}
\begin{table}[htbp]
\caption{Performance evaluation on different subjects}
\begin{center}\label{subject}
\begin{tabular}{c c c c}
\hline
\hline
\textbf{Subjects} &\textbf{Accuracy}  &\textbf{Validation Accuracy} &\textbf{Test Accuracy} \\
&\textbf{(\%)} &\textbf{(\%)} &\textbf{(\%)} \\
\hline
\hline
Subject-1 &96.02 &94.55 &94.431\\
\hline
Subject-2 &93.58 &92.12 &91.09\\
\hline
Subject-3 &96.10 &94.51 &94.12 \\
\hline
Subject-4 &93.01 &90.75 &91.75 \\
\hline
Subject-5 &93.28 &91.47 &91.01 \\
\hline
Subject-6 &95.95 &94.98 &93.02 \\
\hline
Subject-7 &96.72 &94.51 &94.29\\
\hline
Subject-8 &93.44 &90.94 &91.09\\
\hline
Subject-9 &94.02 &91.88 &91.91 \\
\hline
\hline
\end{tabular}
\end{center}
\end{table}

\begin{figure}[htbp]
    \centering
    \includegraphics[width=7in]{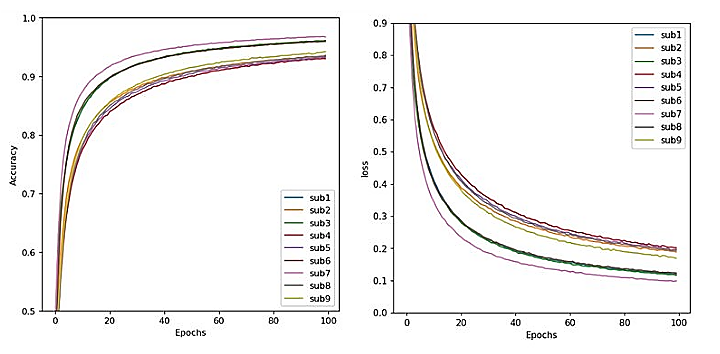}
    \caption{Accuracy V/s Epochs and Loss V/s Epochs for Nine Subjects}
    \label{9sub}
\end{figure}
\begin{figure}[htbp]
    \centering
    \includegraphics[width=7in]{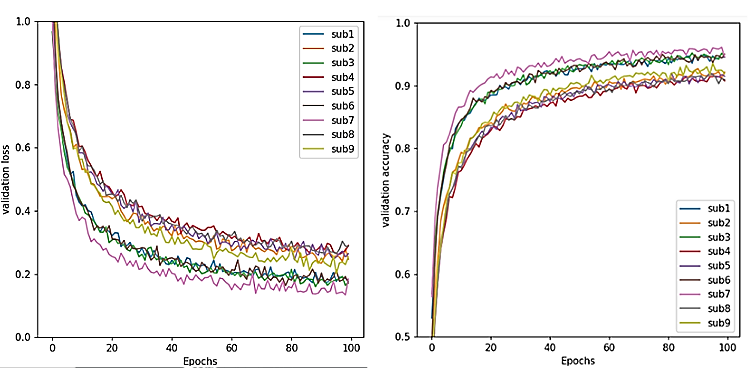}
    \caption{Validation-Accuracy V/s Epochs and Validation-Loss V/s Epochs for Nine Subjects}
    \label{9subv}
\end{figure}

\subsection{Comparison of Proposed model Performance with Deep Learning performance}\vspace{10pt}
The deep-learning architecture was compared with the proposed model. All of the above pre-processing processes were taken into account when developing this method. Only single subject was considered for the evaluation, since the training process was time-consuming. The signals were segmented before being transformed to images and fed into the CNN-architecture as discussed earlier.
\bgroup
\def\arraystretch{0.7}
\begin{table}[htbp]
\caption{Performance Metrics for Deep Learning Architecture}
\begin{center}\label{deep_met}
\begin{tabular}{c c c c c}
\hline
\hline
&\textbf{Precision} &\textbf{Recall} &\textbf{F1-Score}  &\textbf{support}  \\
\hline
\hline
Class 0     &0.8594    &0.8186    &0.8386     &228\\
 \hline
Class 1     &0.86    &0.82    &0.89     &215\\
\hline
Class 2     &0.82    &0.89    &0.80     &233\\
\hline
Class 3     &0.89    &0.87    &0.83     &203\\
\hline
    Accuracy        &           &      &0.83     &879\\
\hline
\hline
\end{tabular}
\end{center}
\end{table}
 Additionally the data was normalised before training and divided the data set into training and validation sets in an 80:20 ratio having batch size 128. The raw data was stored into a pickle file during the training phase. Validation accuracy and accuracy was checked using model-checkpoint callbacks and saved the best performing model in the .h5 file. As a result, the finest model was saved. Hence, no need to train the data-set again and again after completion of this step. The overall-accuracy for deep learning model~\cite{sai2017automated} is 95.41\% and the validation accuracy is 86.76\% for 200 epochs. After comparison of deep learning model with the proposed model. It can be concluded that the proposed model outperformed and consumed less time while training. Precision, Recall, F1-Score was calculated for evaluation for Deep learning Architecture which is shown in Table \ref{deep_met}. 
\subsection{Comparison with State of Art Work}
\bgroup
\def\arraystretch{1.5}
\begin{table}[htbp]
\caption{Comparison with available State-of-the-art }\label{ta}
\begin{center}
\begin{tabular}{l l l l l l}
\hline
\hline
\textbf{Ref.} &\textbf{Year} &\textbf{Feature} &\textbf{Tasks}  &\textbf{Classifier} &\textbf{Decoding} \\
\textbf{No.} & &\textbf{Extraction} & & &\textbf{Performance(\%)}\\
 \hline
 \hline
\cite{7993477} &2016 &Power Spectral Density, &MI Right and,  &SVM &(Class-2) \\
& &Entropy & Left Hand Mt. & &Accuracy-91.25\\
\hline
\cite{7853902} &2016 &MEMD, CSP &MI Right and &SVM &(Class-2)\\
& & & Left Hand Mt. & &Accuracy 83.3 \\
\hline
\cite{8289801} &2017 &AFA, GLCM &MI &SVM,KNN &(Class-2) (SVM+\\
& &DM & &LDA Bayes &AFA)-Acc:89.29\\
\hline
\cite{8298770} &2017 &CSP &MI &SVM &(Class-2)\\
& & & & &Accuracy:95\\
\hline
\cite{mousapour2018novel} &2018 &Connectivity &MI, ME &CNN &(Class-2)96.69\\
& &pattern & & &(Class-4)86\% \\
\hline
\cite{salwa2018classification} &2018 &Maximum &MI &SVM &(Class-2)\\
& &Entropy-HOS & & &Accuracy:79.14\\
\hline
\cite{benzy2019classification} &2019 &Phase Lock &MI &Naive &(Class-2)\\
& &Lock Value & &Bayes &Accuracy:88.2\\
\hline
\cite{chowdhury2019processing} &2019 &Mean,Median,Std &MI Rest,Right and &SVM &(Class-3)\\
& &mode,min,max of CCS &Left Hand Mt. & &Accuracy:78.75\\
\hline
\cite{robinson2019eeg} &2019 &1D to 2D &MI &CNN &(class-2)\\
& & & & &Acc:71.82$(\pm 4.2)$\\
\hline
\cite{shajil2020deep} &2020 &2D- &MI &Inception &(class-2)\\
& &spectrogram & &V3 &Acc:83.79$(\pm3.49)$ \\
\hline
\cite{zhang2019classification} &2020 &Mean,Variance,Entropy &MI, ME &Attention+ &(class-4)\\
&&Pkp,PSD,etc & &LSTM  &Acc:83.2($\pm.12$)\\
\hline
\cite{hossain2020detection} &2020 &17-statistical &MI &SVM &(Class-2)\\
& &features & & &F1-score:68.69\\
\hline
\textbf{Ours} &\textbf{2021} &\textbf{Mean,Var,Skewness} &\textbf{MI,ME, Right} &\textbf{MLP-NN} &\textbf{(Class-4)}\\
& &\textbf{Kurtosis,Area,PSD} &\textbf{\& Left Hand Mt.} &\textbf{CNN+LSTM} &\textbf{1-Acc:  96.02}\\
&&&&&\textbf{9-Acc: 94.47}\\
&&&&&\textbf{CNN- Acc: 86.76}\\
\hline
\hline
\end{tabular}
\end{center}
\label{table1}
\end{table}
\egroup

The performance of the algorithm is compared with previous related work. The results show that our algorithm has best accuracy with more activities classification. 

\section{Conclusions and Future Scope }
A unique algorithm is presented for classifying left and right movements from EEG signals. EEG signals are prone to contaminate and unwanted signals. Successful removal of contamination is done using ICA and a novel approach is developed and implemented. A detailed study of different types of artifacts and also detailed discussed were carried out on how they affect the signal. After successful elimination of artifacts and pre-processing, feature extraction is carried in both domains that is time, as well as frequency domains, and performance in different domains, are compared simultaneously. The algorithm outperformed when both domains feature vectors were included. The intra-subject validation scheme was used to validate on different subjects. Finally, resembled the execution of the proposed model with the deep learning model. From the research, it can be concluded that artifact removal is an essential step. While evaluating the model for EEG signals both time and frequency domain must be considered.

Furthermore, auto-encoders, can be explored to increase the classification performance. The further GANs may explored to generate EEG signals artificially to classify some complex activities. Different feature extraction techniques may be explored using deep learning to evaluate the performance on pre-trained Network. A proper validation strategy for EEG-based categorization has the potential to be developed in the future. There must be a detailed discussion and a consistent technique for removing artifacts. Training requires the development of new inventive schemes that take less time and have a faster reaction time. The future research will primarily concentrate on the real-time implementation of the technology.


\bibliographystyle{unsrt}
\bibliography{mybib_1}

\begin{thebibliography}{10}

\bibitem{zhang2019classification}
Guangyi Zhang, Vandad Davoodnia, Alireza Sepas-Moghaddam, Yaoxue Zhang, and Ali
  Etemad.
\newblock Classification of hand movements from eeg using a deep
  attention-based lstm network.
\newblock {\em IEEE Sensors Journal}, 20(6):3113--3122, 2019.

\bibitem{ABDULKADER2015213}
Sarah~N. Abdulkader, Ayman Atia, and Mostafa-Sami~M. Mostafa.
\newblock Brain computer interfacing: Applications and challenges.
\newblock {\em Egyptian Informatics Journal}, 16(2):213--230, 2015.

\bibitem{pancholi2019improved}
Sidharth Pancholi and Amit~M Joshi.
\newblock Improved classification scheme using fused wavelet packet transform
  based features for intelligent myoelectric prostheses.
\newblock {\em IEEE Transactions on Industrial Electronics}, 67(10):8517--8525,
  2019.

\bibitem{pancholi2019electromyography}
Sidharth Pancholi and Amit~M Joshi.
\newblock Electromyography-based hand gesture recognition system for upper limb
  amputees.
\newblock {\em IEEE Sensors Letters}, 3(3):1--4, 2019.

\bibitem{Pancholi2019EMBC}
Jain P. Varghese A. Joshi Amit~M. Pancholi, S.
\newblock A novel time-domain based feature for emg-pr prosthetic and
  rehabilitation application.
\newblock In {\em In 2019 41st Annual International Conference of the IEEE
  Engineering in Medicine and Biology Society (EMBC)}, pages 5084--5087. IEEE,
  2019.

\bibitem{pancholi2020advanced}
Sidharth Pancholi and Amit~M Joshi.
\newblock Advanced energy kernel-based feature extraction scheme for improved
  emg-pr-based prosthesis control against force variation.
\newblock {\em IEEE Transactions on Cybernetics}, 2020.

\bibitem{jeong2020brain}
Ji-Hoon Jeong, Kyung-Hwan Shim, Dong-Joo Kim, and Seong-Whan Lee.
\newblock Brain-controlled robotic arm system based on multi-directional
  cnn-bilstm network using eeg signals.
\newblock {\em IEEE Transactions on Neural Systems And Rehabilitation
  Engineering}, 28(5):1226--1238, 2020.

\bibitem{jamal2012signal}
Muhammad~Zahak Jamal.
\newblock Signal acquisition using surface emg and circuit design
  considerations for robotic prosthesis.
\newblock {\em Computational Intelligence in Electromyography Analysis-A
  Perspective on Current Applications and Future Challenges}, 18:427--448,
  2012.

\bibitem{pancholi2018portable}
Sidharth Pancholi and Amit~M Joshi.
\newblock Portable emg data acquisition module for upper limb prosthesis
  application.
\newblock {\em IEEE Sensors Journal}, 18(8):3436--3443, 2018.

\bibitem{pancholi2019time}
Sidharth Pancholi and Amit~M Joshi.
\newblock Time derivative moments based feature extraction approach for
  recognition of upper limb motions using emg.
\newblock {\em IEEE Sensors Letters}, 3(4):1--4, 2019.

\bibitem{Pancholi2021iULP}
Amit M.~Joshi Pancholi~Sidharth.
\newblock Intelligent upper-limb prosthetic control (iulp) with novel feature
  extraction method for pattern recognition using emg.
\newblock {\em Journal of Mechanics in Medicine and Biology}, 2021.

\bibitem{mne_ica}
\url{https://mne.tools/stable/auto_tutorials/preprocessing/40_artifact_correction_ica.html}.

\bibitem{Sidharth2021DLPR}
Amit M.~Joshi Pancholi, Sidharth and Deepak Joshi.
\newblock A robust and accurate deep learning based pattern recognition
  framework for upper limb prosthesis using semg.
\newblock {\em arXiv preprint arXiv:2106.02463}, 2021.

\bibitem{7993477}
Shubham Singla, S.~N. Garsha, and Somsirsa Chatterjee.
\newblock Characterization of classifier performance on left and right limb
  motor imagery using support vector machine classification of eeg signal for
  left and right limb movement.
\newblock In {\em 2016 5th International Conference on Wireless Networks and
  Embedded Systems (WECON)}, pages 1--4, 2016.

\bibitem{mousapour2018novel}
Leila Mousapour, Fateme Agah, Soorena Salari, and Marzieh Zare.
\newblock A novel approach to classify motor-imagery eeg with convolutional
  neural network using network measures.
\newblock In {\em 2018 4th Iranian Conference on Signal Processing and
  Intelligent Systems (ICSPIS)}, pages 43--47. IEEE, 2018.

\bibitem{8298770}
Su~Jing and Xie Yun.
\newblock Off-line analysis of motor imagery electroencephalogram.
\newblock In {\em 2017 First International Conference on Electronics
  Instrumentation Information Systems (EIIS)}, pages 1--6, 2017.

\bibitem{salwa2018classification}
LAGDALI Salwa et~al.
\newblock Classification of left and right hand movement from eeg signals by
  hos features.
\newblock In {\em 2018 9th International Symposium on Signal, Image, Video and
  Communications (ISIVC)}, pages 314--317. IEEE, 2018.

\bibitem{benzy2019classification}
VK~Benzy and AP~Vinod.
\newblock Classification of motor imagery hand movement directions from eeg
  extracted phase locking value features for brain computer interfaces.
\newblock In {\em TENCON 2019-2019 IEEE Region 10 Conference (TENCON)}, pages
  2315--2319. IEEE, 2019.

\bibitem{chowdhury2019processing}
Md~Abu~Shahid Chowdhury and Dabasish~Kumar Saha.
\newblock Processing of motor imagery eeg signals for controlling the opening
  and the closing of artificial hand.
\newblock In {\em 2019 4th International Conference on Electrical Information
  and Communication Technology (EICT)}, pages 1--5. IEEE, 2019.

\bibitem{9219507}
Hasan Polat and Mehmet~Siraç Özerdem.
\newblock Automatic detection of cursor movements from the eeg signals via deep
  learning approach.
\newblock In {\em 2020 5th International Conference on Computer Science and
  Engineering (UBMK)}, pages 327--332, 2020.

\bibitem{hossain2020detection}
Muhammad~Yeamim Hossain and ABMSU Doulah.
\newblock Detection of motor imagery (mi) event in electroencephalogram (eeg)
  signals using artificial intelligence technique.
\newblock In {\em 2020 IEEE East-West Design \& Test Symposium (EWDTS)}, pages
  1--6. IEEE, 2020.

\bibitem{sanei2013adaptive}
Saeid Sanei.
\newblock {\em Adaptive processing of brain signals}.
\newblock John Wiley \& Sons, 2013.

\bibitem{bansal2019eeg}
Dipali Bansal and Rashima Mahajan.
\newblock {\em EEG-Based Brain-Computer Interfaces: Cognitive Analysis and
  Control Applications}.
\newblock Academic Press, 2019.

\bibitem{de2009handbook}
Michelle De~Haan and Megan~R Gunnar.
\newblock {\em Handbook of developmental social neuroscience}.
\newblock Guilford Press, 2009.

\bibitem{zhang2020arder}
Chenbei Zhang, Yong Lian, and Guoxing Wang.
\newblock Arder: An automatic eeg artifacts detection and removal system.
\newblock In {\em 2020 27th IEEE International Conference on Electronics,
  Circuits and Systems (ICECS)}, pages 1--2. IEEE, 2020.

\bibitem{mashhadi2020deep}
Najmeh Mashhadi, Abolfazl~Zargari Khuzani, Morteza Heidari, and Donya
  Khaledyan.
\newblock Deep learning denoising for eog artifacts removal from eeg signals.
\newblock {\em arXiv preprint arXiv:2009.08809}, 2020.

\bibitem{sai2017automated}
Chong~Yeh Sai, Norrima Mokhtar, Hamzah Arof, Paul Cumming, and Masahiro
  Iwahashi.
\newblock Automated classification and removal of eeg artifacts with svm and
  wavelet-ica.
\newblock {\em IEEE journal of biomedical and health informatics},
  22(3):664--670, 2017.

\bibitem{4536072}
Jurgen Dammers, Michael Schiek, Frank Boers, Carmen Silex, Mikhail Zvyagintsev,
  Uwe Pietrzyk, and Klaus Mathiak.
\newblock Integration of amplitude and phase statistics for complete artifact
  removal in independent components of neuromagnetic recordings.
\newblock {\em IEEE Transactions on Biomedical Engineering}, 55(10):2353--2362,
  2008.

\bibitem{7319296}
Irene Winkler, Stefan Debener, Klaus-Robert Müller, and Michael Tangermann.
\newblock On the influence of high-pass filtering on ica-based artifact
  reduction in eeg-erp.
\newblock In {\em 2015 37th Annual International Conference of the IEEE
  Engineering in Medicine and Biology Society (EMBC)}, pages 4101--4105, 2015.

\bibitem{jain2020iglu}
Prateek Jain, Amit~M Joshi, Navneet Agrawal, and Saraju Mohanty.
\newblock iglu 2.0: A new non-invasive, accurate serum glucometer for smart
  healthcare.
\newblock {\em arXiv preprint arXiv:2001.09182}, 2020.

\bibitem{joshi2020iglu2.0}
Amit~M Joshi, Prateek Jain, Saraju~P Mohanty, and Navneet Agrawal.
\newblock iglu 2.0: A new wearable for accurate non-invasive continuous serum
  glucose measurement in iomt framework.
\newblock {\em IEEE Transactions on Consumer Electronics}, 66(4):327--335,
  2020.

\bibitem{kansara2018visual}
Pankti Kansara.
\newblock Visual question answering.
\newblock 2018.

\bibitem{8922820}
Prateek Jain, Amit~M. Joshi, and Saraju~P. Mohanty.
\newblock iglu: An intelligent device for accurate noninvasive blood
  glucose-level monitoring in smart healthcare.
\newblock {\em IEEE Consumer Electronics Magazine}, 9(1):35--42, 2020.

\bibitem{sharma2021dephnn}
Geetanjali Sharma, Abhishek Parashar, and Amit~M Joshi.
\newblock Dephnn: A novel hybrid neural network for electroencephalogram
  (eeg)-based screening of depression.
\newblock {\em Biomedical Signal Processing and Control}, 66:102393, 2021.

\bibitem{kingma2014adam}
Diederik~P Kingma and Jimmy Ba.
\newblock Adam: A method for stochastic optimization.
\newblock {\em arXiv preprint arXiv:1412.6980}, 2014.

\bibitem{reddi2019convergence}
Sashank~J Reddi, Satyen Kale, and Sanjiv Kumar.
\newblock On the convergence of adam and beyond.
\newblock {\em arXiv preprint arXiv:1904.09237}, 2019.

\bibitem{fahimi2019towards}
Fatemeh Fahimi, Zhuo Zhang, Wooi~Boon Goh, Kai~Keng Ang, and Cuntai Guan.
\newblock Towards eeg generation using gans for bci applications.
\newblock In {\em 2019 IEEE EMBS International Conference on Biomedical \&
  Health Informatics (BHI)}, pages 1--4. IEEE, 2019.

\bibitem{SUN20081663}
Shiliang Sun, Changshui Zhang, and Yue Lu.
\newblock The random electrode selection ensemble for eeg signal
  classification.
\newblock {\em Pattern Recognition}, 41(5):1663--1675, 2008.

\bibitem{ward2019student}
Jamie Ward.
\newblock {\em The student’s guide to cognitive neuroscience}.
\newblock Routledge, 2019.

\bibitem{EEGLAB}
Oostenveld R Onton~J Delorme~A, Palmer~J.
\newblock Tutorials for understanding eeg siganls, 2021.

\bibitem{dammers2008integration}
Jurgen Dammers, Michael Schiek, Frank Boers, Carmen Silex, Mikhail Zvyagintsev,
  Uwe Pietrzyk, and Klaus Mathiak.
\newblock Integration of amplitude and phase statistics for complete artifact
  removal in independent components of neuromagnetic recordings.
\newblock {\em IEEE transactions on biomedical engineering}, 55(10):2353--2362,
  2008.

\bibitem{kaur2017detection}
Jaspreet Kaur, Sidharth Pancholi, and Amit~M Joshi.
\newblock Detection of lung cancer with the fusion of computed tomography and
  positron emission tomography.
\newblock In {\em International Conference on Next Generation Computing
  Technologies}, pages 944--955. Springer, 2017.

\bibitem{boonyakitanont2020review}
Poomipat Boonyakitanont, Apiwat Lek-Uthai, Krisnachai Chomtho, and Jitkomut
  Songsiri.
\newblock A review of feature extraction and performance evaluation in
  epileptic seizure detection using eeg.
\newblock {\em Biomedical Signal Processing and Control}, 57:101702, 2020.

\bibitem{jain2019iglu}
Prateek Jain, Amit~M Joshi, and Saraju~P Mohanty.
\newblock iglu 1.0: An accurate non-invasive near-infrared dual short
  wavelengths spectroscopy based glucometer for smart healthcare.
\newblock {\em arXiv preprint arXiv:1911.04471}, 2019.

\bibitem{li2020electroencephalography}
Zheng Li.
\newblock Electroencephalography signal analysis and classification based on
  deep learning.
\newblock In {\em 2020 5th International Conference on Information Science,
  Computer Technology and Transportation (ISCTT)}, pages 119--125. IEEE, 2020.

\bibitem{pancholi2021robust}
Sidharth Pancholi, Amit~M Joshi, and Deepak Joshi.
\newblock A robust and accurate deep learning based pattern recognition
  framework for upper limb prosthesis using semg.
\newblock {\em arXiv preprint arXiv:2106.02463}, 2021.

\bibitem{tadalagi2021autodep}
Manjunath Tadalagi and Amit~M Joshi.
\newblock Autodep: automatic depression detection using facial expressions
  based on linear binary pattern descriptor.
\newblock {\em Medical \& Biological Engineering \& Computing}, pages 1--16,
  2021.

\bibitem{7853902}
Sanjoy~Kumar Saha and Md.~Sujan Ali.
\newblock Data adaptive filtering approach to improve the classification
  accuracy of motor imagery for bci.
\newblock In {\em 2016 9th International Conference on Electrical and Computer
  Engineering (ICECE)}, pages 247--250, 2016.

\bibitem{8289801}
Muhammed Al-Suify, Walid Al-Atabany, and Mohamed A.~A. Eldosoky.
\newblock Classification of right and left hand movement using nonlinear
  analysis.
\newblock In {\em 2017 13th International Computer Engineering Conference
  (ICENCO)}, pages 282--285, 2017.

\bibitem{robinson2019eeg}
Neethu Robinson, Seong-Whan Lee, and Cuntai Guan.
\newblock Eeg representation in deep convolutional neural networks for
  classification of motor imagery.
\newblock In {\em 2019 IEEE International Conference on Systems, Man and
  Cybernetics (SMC)}, pages 1322--1326. IEEE, 2019.

\bibitem{shajil2020deep}
Nijisha Shajil, M~Sasikala, and AM~Arunnagiri.
\newblock Deep learning classification of two-class motor imagery eeg signals
  using transfer learning.
\newblock In {\em 2020 International Conference on e-Health and Bioengineering
  (EHB)}, pages 1--4. IEEE, 2020.

\end{thebibliography}

\pagebreak


\section*{Authors' Biographies}



\begin{minipage}[htbp]{\columnwidth}
	\begin{wrapfigure}{l}{1.00in}
		\vspace{-0.3cm}
		\includegraphics[width=1.0in,keepaspectratio]{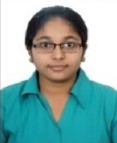}
		\vspace{-0.5cm}
	\end{wrapfigure}
	\noindent
	\textbf{Pranali Kokate} is a M.Tech student at Malaviya National Institute of Technology (MNIT). She has one year of industrial experience in Embedded Systems. Her current research interests include Machine Learning, Brain-Computer Interface, Deep Learning, Embedded Systems, Embedded Linux and Industrial Internet of Things.
	\end{minipage}

\vspace{1.8cm}

\begin{minipage}[htbp]{\columnwidth}
	\begin{wrapfigure}{l}{1.0in}
		\vspace{-0.4cm}
		\includegraphics[width=1.0in,keepaspectratio]{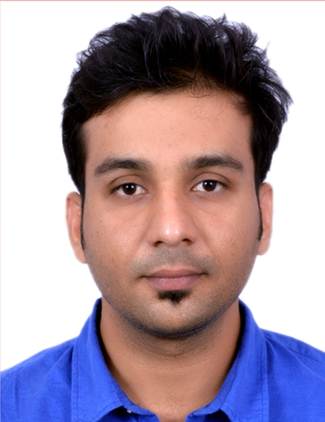}
		\vspace{-0.5cm}
	\end{wrapfigure}
	\noindent
		\textbf{Sidharth Pancholi} (GS'18) received the Masters from Thapar University, Patiala, India, in 2016. He has
completed his PhD work with Malaviya National
Institute of Technology, Jaipur, India. Currently,
he is working as a Research Associate with the
Indian Institute of Technology, Delhi, India. He
has worked as a reviewer of technical journals
such as IEEE Transactions/ journals and served
as Technical Programme Committee member for
IEEE conferences. He also received CSIR Travel
fellowship and CSSTDS Travel fellowship to attend
IEEE Conference EMBC 2019. He has two years of industry experience.
His current research interests include biomedical signal processing, neural
rehabilitation, Brain computer interface, Human-machine interface, prosthetic
device development, and embedded systems.
\end{minipage}

\vspace{1.8cm}

\begin{minipage}[htbp]{\columnwidth}
	\begin{wrapfigure}{l}{1.0in}
		\vspace{-0.4cm}
		\includegraphics[width=1.0in,keepaspectratio]{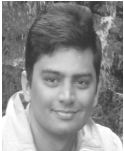}
		\vspace{-0.5cm}
	\end{wrapfigure}
	\noindent
	\textbf{Amit M. Joshi} (M'08) has completed his M.Tech (by research) in 2009 and obtained Doctoral of Philosophy degree (Ph.D) from National Institute of Technology, Surat in August 2015. He is currently an Assistant Professor at National Institute of Technology, Jaipur since July 2013. His area of specialization is Biomedical signal processing, Smart healthcare, VLSI DSP Systems and embedded system design. He has published six book chapters and also published 50+ research articles in peer reviewed international journals/conferences. He has served as a reviewer of technical journals such as IEEE Transactions, Springer, Elsevier and also served as Technical Programme Committee member for IEEE conferences. He also received UGC Travel fellowship, SERB DST Travel grant  and CSIR Travel fellowship to attend IEEE Conferences in VLSI and Embedded System. He has served session chair at various IEEE Conferences like TENCON -2016, iSES-2018, ICCIC-14. He has already supervised 18 M.Tech projects and 14 B.Tech projects in the field of VLSI and Embedded Systems and VLSI DSP systems. He is currently supervising six  Ph.D. students.
\end{minipage}


\end{document}